\documentclass{fmcad}
\ifCLASSOPTIONcompsoc
  \usepackage[nocompress]{cite}
\else
  \usepackage{cite}
\fi
\ifCLASSINFOpdf
\else
\fi

\usepackage{color}
\usepackage{url}
\usepackage{xspace}
\usepackage{amsmath}
\usepackage{amsfonts}
\usepackage{listings}
\usepackage{graphicx}
\usepackage{algpseudocode}
\usepackage{algorithm}
\usepackage{mathtools}
\usepackage[mathscr]{euscript}
\usepackage{comment}
\usepackage{caption}
\usepackage{subcaption}
\usepackage{bussproofs}
\usepackage{mathpartir}
\usepackage{enumitem}
\usepackage{times}
\usepackage{lipsum}
\usepackage{float}

\usepackage{amssymb}
\usepackage{amsthm}

\algnewcommand\algorithmicforeach{\textbf{for each}}
\algdef{S}[FOR]{ForEach}[1]{\algorithmicforeach\ #1\ \algorithmicdo}

\algnewcommand\algorithmicinput{\textbf{Input:}}
\algnewcommand\algorithmicoutput{\textbf{Output:}}
\algnewcommand\Input{\item[\algorithmicinput]}
\algnewcommand\Output{\item[\algorithmicoutput]}

\algnewcommand\algorithmicproc{\textbf{Procedure}}
\algnewcommand\procedure{\item[\algorithmicproc]}

\newcommand{\nt}[1]{\textcolor[RGB]{148,0,211}{\textit{#1} -NT}\xspace}

\newcommand{\define}[1]{\textsl{#1}}
\newcommand{\Mo}{\mathbf{I}}
\newcommand{\Model}{\mathcal{I}}
\newcommand{\T}{\mathcal{T}}
\newcommand{\Prime}{\mathit{prime}}
\newcommand{\unroll}{\mathit{unroll}}
\newcommand{\config}{\mathit{conf}}
\newcommand{\cfg}{\mathit{cfg}}
\newcommand{\bool}{\mathit{Bool}}
\newcommand{\ite}{\mathit{ite}}
\newcommand{\true}{\mathit{True}}
\newcommand{\dimensionality}{\mathit{dim}}
\newcommand{\ranges}{\mathit{ranges}}
\newcommand{\strides}{\mathit{strides}}
\newcommand{\startingaddr}{\mathit{offset}}
\newcommand{\vals}{\mathit{vals}}

\newcommand{\pono}{\texttt{Pono}\xspace}
\newcommand{\smtswitch}{\texttt{Smt-Switch}\xspace}
\newcommand{\btor}{\texttt{Boolector}\xspace}
\newcommand{\yosys}{\texttt{Yosys}\xspace}
\newcommand{\vect}[1]{\boldsymbol{#1}}

\newtheorem{theorem}{Theorem}[section]
\newtheorem{example}{Example}

\usepackage{graphicx}

\begin{document}
%
% paper title
% Titles are generally capitalized except for words such as a, an, and, as,
% at, but, by, for, in, nor, of, on, or, the, to and up, which are usually
% not capitalized unless they are the first or last word of the title.
% Linebreaks \\ can be used within to get better formatting as desired.
% Do not put math or special symbols in the title.
\title{Automating System Configuration}

% author names and affiliations
% use a multiple column layout for up to three different
% affiliations

\author{\IEEEauthorblockN{Nestan Tsiskaridze, Maxwell Strange, Makai Mann, Kavya Sreedhar, Qiaoyi Liu, Mark Horowitz, Clark Barrett}
\IEEEauthorblockA{Stanford University, Stanford, CA 94305, USA\\
E-mail: \{nestan, mstrange, makaim, skavya, joeyliu\}@stanford.edu,  horowitz@ee.stanford.edu, barrett@cs.stanford.edu}
}

% conference papers do not typically use \thanks and this command
% is locked out in conference mode. If really needed, such as for
% the acknowledgment of grants, issue a \IEEEoverridecommandlockouts
% after \documentclass

% for over three affiliations, or if they all won't fit within the width
% of the page (and note that there is less available width in this regard for
% compsoc conferences compared to traditional conferences), use this
% alternative format:
% 
%\author{\IEEEauthorblockN{Michael Shell\IEEEauthorrefmark{1},
%Homer Simpson\IEEEauthorrefmark{2},
%James Kirk\IEEEauthorrefmark{3}, 
%Montgomery Scott\IEEEauthorrefmark{3} and
%Eldon Tyrell\IEEEauthorrefmark{4}}
%\IEEEauthorblockA{\IEEEauthorrefmark{1}School of Electrical and Computer Engineering\\
%Georgia Institute of Technology,
%Atlanta, Georgia 30332--0250\\ Email: see http://www.michaelshell.org/contact.html}
%\IEEEauthorblockA{\IEEEauthorrefmark{2}Twentieth Century Fox, Springfield, USA\\
%Email: homer@thesimpsons.com}
%\IEEEauthorblockA{\IEEEauthorrefmark{3}Starfleet Academy, San Francisco, California 96678-2391\\
%Telephone: (800) 555--1212, Fax: (888) 555--1212}
%\IEEEauthorblockA{\IEEEauthorrefmark{4}Tyrell Inc., 123 Replicant Street, Los Angeles, California 90210--4321}}

% use for special paper notices
%\IEEEspecialpapernotice{(Invited Paper)}

% make the title area
\maketitle

% As a general rule, do not put math, special symbols or citations
% in the abstract
\begin{abstract}
%The abstract should briefly summarize the contents of the paper in 
%150--250 words.
The increasing complexity of modern configurable systems makes it critical to improve the level of automation in the process of system configuration. Such automation can also improve the agility of the development cycle, allowing for rapid and automated integration of decoupled workflows.  In this paper, we present a new framework for automated configuration of systems representable as state machines. The framework leverages model checking and satisfiability modulo theories (SMT) and can be applied to any application domain representable using SMT formulas.   Our approach can also be applied modularly, improving its scalability.  Furthermore, we show how optimization can be used to produce configurations that are best according to some metric and also more likely to be understandable to humans.  We showcase this framework and its flexibility by using it to configure a CGRA memory tile for various image processing applications.  

%\keywords{Automated system configuration \and model checking \and Satisfiability Modulo Theories.}
\end{abstract}
% no keywords

% For peer review papers, you can put extra information on the cover
% page as needed:
% \ifCLASSOPTIONpeerreview
% \begin{center} \bfseries EDICS Category: 3-BBND \end{center}
% \fi
%
% For peerreview papers, this IEEEtran command inserts a page break and
% creates the second title. It will be ignored for other modes.
\IEEEpeerreviewmaketitle

\section{Introduction}

In systems engineering, the \emph{system configuration} problem arises when systems are parameterized to increase their flexibility or functionality.  It refers to the problem of choosing the appropriate parameter values for the context or application in which the system will be used. Most hardware and software systems, including hardware IPs, operating systems, networks, servers, and data centers, require some degree of configuration. 
The need for configuration also often arises when integrating decoupled parts of a system, including integrating software and hardware.

The difficulty of the system configuration problem has been gradually growing as systems increase in scale and complexity.  In particular, in an effort to make designs more widely applicable and re-usable, there has been an increasing use of hardware that is configurable, not only at design time or setup time, but even during normal operation.
Manual configuration of such systems is error-prone and may even be impossible, depending on how frequently the systems need to be reconfigured. 

Automation of the configuration problem can also be beneficial during the system design process.  In particular, it obviates the need for new hand-coded configuration files every time some configurable component changes.  Increased automation of such steps supports a move towards more agile design processes.  Agile approaches typically require the ability to rapidly and (largely) automatically integrate changing parts of a system while continuously maintaining correct end-to-end functionality.  Having design blocks that are flexibly configurable aids this effort, as does the ability to automate the configuration.

A potential disadvantage of automated configuration is that it could lead to an increase in the opacity of the overall system.  Hand-written configurations can be documented and explained to allow for easier understandability and maintainability.  Thus, an additional goal when automating configuration should be to produce results that are comprehensible to humans and that can be easily reviewed and maintained.

In this paper, we present a general framework for automated system configuration.  It provides a flexible approach for solving the configuration problem for systems composed of software, hardware, or both. The systems are modeled using transition systems, where transition formulas can use the full expressive power of SMT-LIB~\cite{smt-lib}, the language used by satisfiability modulo theories (SMT)~\cite{smt} solvers.
The framework provides a systematic approach to facilitate fully automated or automation-guided system configuration. It is well-suited for both stand-alone designs and for designs with multiple configurable parts.  For the latter, it is especially useful during system integration and rapid development.

The main contributions of this paper are:

\begin{itemize}
    \item We introduce a ``programming by example'' approach for formalizing common input-output specifications. In an exact formulation of the configuration problem, the input-output specification might need to universally quantify over the input variables. Our approach avoids the need for quantifiers.
    \item We propose a new modular approach for configuration finding in a general SMT setting that makes use of abduction. \item We show how to leverage optimization to obtain human-readable configurations.
    \item We present a case study---automated configuration of a memory tile in the context of an agile hardware design project targeting image processing applications.
\end{itemize}

The remainder of the paper is organized as follows. Section~\ref{section:background} presents background and notation. Section~\ref{section:framework} formalizes the configuration solving problem and introduces our framework, including some extensions and limitations. In Section~\ref{section:optimization}, we show how optimization techniques can be integrated into the approach, both for the purpose of improving performance as well as for improving human readability, and we discuss a few additional extensions of the framework. 
In Section~\ref{section:CaseStudy} we present a case study, giving the details of a specific system design and showing how our framework can be applied.  Experimental results for this case study are then reported in Section~\ref{section:evaluation}. We survey the related work in Section~\ref{section:related} and conclude in Section~\ref{section:conclusion}.

\section{Background}
\label{section:background}

We assume the standard many-sorted first-order logic setting with the usual
notions of signature, term, formula, and interpretation.
A \define{theory} is a pair $\T = (\Sigma, \Mo)$ where
$\Sigma$ is a signature and  $\Mo$ is a class of $\Sigma$-interpretations, i.e.,
the \define{models} of $\T$. 
A $\Sigma$-formula $\varphi$ is
\define{satisfiable} (resp., \define{unsatisfiable}) \define{in $\T$}
if it is satisfied by some (resp., no) interpretation in $\Mo$. 
We define $\models_\T$ over $\Sigma$-formulas: if $\varphi$ and $\psi$ are $\Sigma$-formulas, then $\varphi \models_\T \psi$ if all interpretations which satisfy $\varphi$ also satisfy $\psi$.  In this case, we also call $\varphi$ an \define{abduct} of $\psi$ under $\T$.  For generality, we assume
an arbitrary but fixed background theory $\T$ (which could be a combination of theories) with signature $\Sigma$ and an infinite set $\mathcal{X}$ of variables. We will assume that all terms
and formulas are $\Sigma$-terms and $\Sigma$-formulas whose free variables are in $\mathcal{X}$, that entailment is
entailment modulo $\T$, and that interpretations are $\T$-interpretations that assign every variable in $\mathcal{X}$. 

Given an interpretation $\Model$, a variable assignment $s$ over a set of
variables $V$ is a mapping that assigns each variable $v\in V$ of sort $\sigma$
to an element of $\sigma^{\Model}$, denoted $v^s$.  The assignment over $V$ \emph{induced} by an interpretation $\Model$  (i.e., the assignment that maps each variable in $V$ to its interpretation in $\Model$) is denoted $\Model^V$. 
The assignment $s$ restricted to the domain $U \subseteq V$ is denoted by $s^U$. 
We write $\Model[s]$  for the interpretation
that is equivalent to $\Model$ except that each variable $v\in V$ is mapped to $v^s$.  We write $f \circ g$ for functional composition, i.e.,
$f \circ g (x) = f(g(x))$.

\vspace{1ex}
\noindent
{\bf Satisfiability Modulo Theories (SMT).}
Satisfiability Modulo Theories~\cite{smt} is an extension of the Boolean satisfiability (SAT) problem to satisfiability in first-order theories. SMT solvers combine the Boolean reasoning of a SAT solver with specialized theory solvers to check satisfiability of many-sorted first-order logic formulas. Some examples of commonly supported theories are: fixed-width bit-vectors, uninterpreted functions, linear arithmetic, and arrays. In our case study, we utilize fixed-width bit-vectors for modeling a hardware design.

\vspace{1ex}
\noindent
{\bf Symbolic Transition Systems.}
 
A symbolic transition system (STS) $\mathcal{S}$ is a tuple $\mathcal{S} := \langle V, I, T
\rangle $, where $V$ is a finite set of state variables (possibly of different sorts), $I(V)$ is a formula
denoting the initial states of the system, and $T(V, V')$ is a formula
expressing a transition relation, with $V'$ defined as follows.
Let $\Prime$ be a bijection that maps each variable $v\in V$ to a new variable (not in $V$) $v'$ of the same sort. $V'$ is the codomain of $\Prime$.

A state $s$ of $\mathcal{S}$ is a variable assignment over $V$.  A sequence of states is called a \emph{path}.
An
\emph{execution} of $\mathcal{S}$ of length $k$ is a pair $\langle
\Model,\pi\rangle$, where $\Model$ is an interpretation and $\pi := s_0, s_1,
\ldots, s_{k-1}$ is a \emph{path} such that
$\Model[s_0] \models I(V)$ and $\Model[s_i][s_{i+1} \circ \Prime^{-1}] \models
T(V, V')$ for all $0 \leq i < k - 1$.

\vspace{1ex}
\noindent
{\bf Unrolling and Bounded Model Checking.}
 
An \emph{unrolling} of length $k$ of a symbolic transition system is a formula that captures an execution of length $k$ by creating copies of the transition relation.  This is accomplished by introducing fresh copies of every state variable for each state in the execution path.  We use $V@i$ to
denote the set of variables obtained by replacing each variable $v\in V$ with a new variable called $v@i$ of the same sort. We refer to these as \emph{timed} variables. Given an STS $\mathcal{S}$, let $\unroll(\mathcal{S},k) = I(V@0) \wedge \bigwedge_{0 \leq i < k} T(V@i, V@(i+1))$.

Bounded model checking (BMC)~\cite{bmc} is an unrolling-based symbolic model checking approach. Let $P(V)$ be a formula representing a desired property of a symbolic transition system.  BMC creates an unrolled transition system and adds an additional constraint that the property is violated at time $k$.
The BMC formula at bound $k$ is thus: $\unroll(\mathcal{S},k) \wedge \neg P(V@k)$. A typical approach for BMC starts with $k=0$ and incrementally increases it if no counterexample is found at the current bound. A satisfiable BMC formula can easily be converted into an execution that violates the property.

\vspace{1ex}
\noindent
{\bf Optimization.}
An \emph{optimization problem} $\mathcal{OP}$ is a tuple $\langle t, A, \preccurlyeq, \phi, \mathcal{O} \rangle$ where:
\begin{itemize}
    \item $t$ is an \emph{objective term} to optimize of sort $\sigma$; 
    \item $A$ is a set and $\preccurlyeq$ is a total order over $A$.
    \item $\phi$ is a formula to satisfy; and
    \item $\mathcal{O}\! \in\! \{\mathit{min},\!\mathit{max}\}$ is the optimization objective. 
\end{itemize}

\noindent
$\Model$ is a solution to $\mathcal{OP}$ if $\sigma^{\Model} = A$, $\Model \models \phi$, and for any $\Model'$, such that $\sigma^{\Model'}=A$ and $\Model' \models \phi$:\\[-.2cm] 
\[{(\mathcal{O} \! = \! \mathit{min} \! \to t^{\Model} \! \preccurlyeq  t^{\Model'}) \; \wedge \; (\mathcal{O} \! = \! max \! \to \! t^{\Model'} \! \preccurlyeq  t^{\Model}).}\]

A \emph{multi-objective optimization problem} $\mathcal{MOP}$ is a finite sequence of optimization problems $\{\mathcal{OP}_1,\dots,\mathcal{OP}_n\}$ over the same formula $\phi$, where $\mathcal{OP}_i := \langle t_i, A_i, \preccurlyeq_i, \phi, \mathcal{O}_i\rangle$ and $t_i$ is of sort $\sigma_i$ for $i \in [1,n]$. $\Model$ is a solution to $\mathcal{MOP}$ if $\sigma_i^{\Model}=A_i$, $\Model \models \phi$, and for any $\Model'$, such that $\sigma_i^{\Model'}=A_i$ and $\Model' \models \phi$, either:
\begin{itemize}
\item[(i)] $t_i^{\Model} = t_i^{\Model'}$ for all $i \in [1,n]$; or
\item[(ii)] for some $j \! \in \! [1,n]$,
        $t_i^{\Model} = t_i^{\Model'}$ for all $i\in[1,j)$, and\\[-.25cm]
        \[(\mathcal{O}_{j}  =  \mathit{min}  \to  t_j^{\Model} \prec_{j}  t_j^{\Model'}) \wedge (\mathcal{O}_{j}  =  \mathit{max}  \to  t_j^{\Model'}  \prec_{j}  t_j^{\Model}),\]
\end{itemize}
where $\prec$ is the strict total order associated with $\preccurlyeq$.

\section{Configuration Solving Framework}
\label{section:framework}

In this section, we formalize the configuration problem and introduce our automated framework for solving it.  
We also describe how to improve scalability using a modular approach.

\subsection{Problem Formalization}
\label{subsec:cp}
Suppose we have a configurable system that we want to use in a particular application context.  We assume the application context can precisely define an input/output relationship that it expects the system to adhere to.  The \emph{configuration finding problem} is then: given a system $S$ and an application-supplied input-output relationship $P$ for $S$, find a configuration $\mathcal{C}$ for $S$ such that $S$ satisfies $P$ with configuration $\mathcal{C}$.  In this paper, we assume that $P$ specifies behavior for only a finite number of steps. The rationale is that for many configurable systems, a segment of a desired execution is sufficient to partially (or fully) determine what the configuration should be. This is the case for the systems we target and for the case study we describe later. More general specifications are an important direction for future work.

\begin{figure}[t]
    \centering
    \includegraphics[scale=.4]{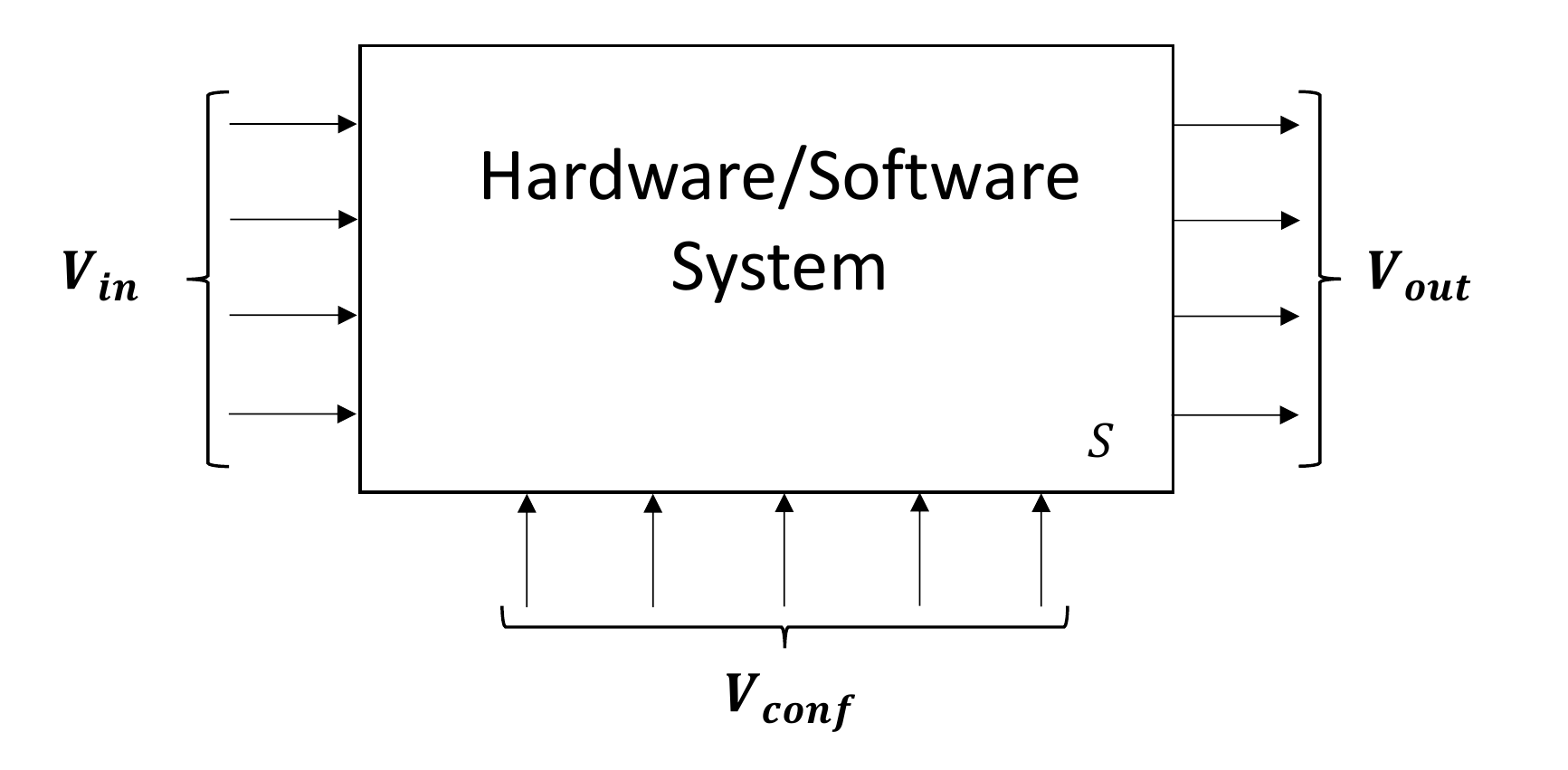}
    \caption{\small{Formal system model.}}
    \label{fig:sys}
\end{figure}

Formally, a configuration problem $\mathcal{CP}$ is a tuple $\langle \mathcal{S}, k,  V_{\mathsf{in}}, V_{\mathsf{out}}, V_{\mathsf{conf}}, P \rangle$ 
where:
\begin{itemize}
    \item $\mathcal{S}:= \langle V, I, T
\rangle $  is a symbolic transition system representing a configurable system $S$, as in Figure~\ref{fig:sys};

\item $k$ is the number of transitions over which the input-output specification will be defined;

\item $V_{\mathsf{in}}, V_{\mathsf{out}}, V_{\mathsf{conf}}$ are three distinguished subsets of the state variables $V$ of $\mathcal{S}$; $V_{\mathsf{in}}$ contains \emph{input} variables (input variables do not appear in $I(V)$, and their primed versions do not appear in $T$); $V_{\mathsf{out}}$ contains \emph{output} variables; and $V_{\mathsf{conf}}\not=\emptyset$ contains the \emph{configuration} variables;  pairwise intersections of these sets may either be empty or non-empty, and $V$ may contain variables that are not in any of these sets; and

\item $P$ is \emph{an input-output property}, or \emph{an input-output specification}, a formula capturing an input-output relationship for $k$ transitions: $P(V_{\mathsf{in}}@0,\dots,V_{\mathsf{in}}@(k-1), V_{\mathsf{out}}@0,\dots,V_{\mathsf{out}}@k)$; a common specification we will use is a set of exact values on input and output variables at each transition: $\bigwedge_{0 \leq i < k} V_{\mathsf{in}}@i = c^i_{in} \wedge \bigwedge_{0 \leq i \leq k} V_{\mathsf{out}}@i = c^i_{out}$.

\end{itemize}

\noindent
A \emph{configuration} $\mathcal{C}$ is defined as an assignment to the variables in $V_{\mathsf{conf}}$.

In this paper, we assume the configuration variables $V_{\mathsf{conf}}$ remain unchanged once configured (a reasonable assumption for many systems, including the one in the case study we present in Section~\ref{section:CaseStudy}). We enforce this by explicitly adding an additional \emph{configuration constancy constraint}: $\config(V_{\mathsf{conf}},k) = \bigwedge_{0 \leq i < k} V_{\mathsf{conf}}@(i+1)=V_{\mathsf{conf}}@i$.
The configuration finding problem then reduces to checking the satisfiability of the \emph{configuration formula}:
%\begin{center}
\begin{multline}\label{F}
\phi(\mathcal{CP}) = \unroll(\mathcal{S},k) \wedge
\config(V_{\mathsf{conf}},k)\ \wedge\\
P(V_{\mathsf{in}}@0,\dots,V_{\mathsf{in}}@(k-1), V_{\mathsf{out}}@0,\dots,V_{\mathsf{out}}@k) 
\end{multline}
%\end{center}
\noindent
A configuration $\mathcal{C}$ is \emph{correct} for $\mathcal{CP}$ if there exists an interpretation $\Model$ such that $\Model \models \phi$ and $\mathcal{C}=\Model^{V_{\mathsf{conf}}}$.

\begin{example}(\textbf{simple ALU})

Let $\mathcal{S} := \langle \{x:int, a:int, \cfg:\bool\}, x = 0, x' = \ite(\cfg, x + a, x - a) \rangle$ be a transition system in a configuration finding problem, where $V_{\mathsf{in}} = \{a\}$, $V_{\mathsf{out}} = \{x\}$, $V_{\mathsf{conf}} = \{\cfg\}$, and $ite$ is the if-then-else operator. There are two ways to configure $\mathcal{S}$: as a system that always adds the current input to the current state, or as a system that always subtracts the current input from the current state. Let us consider two instances of an input-output relation for $k = 2$:
\begin{enumerate}
    \item $P_1(a@0, a@1, x@0,\dots, x@2) = a@0 = 1 \wedge a@1 = 1 \wedge x@0 = 0 \wedge x@1 = 1 \wedge x@2 = 2$. We are interested in whether there exists a value of $\cfg$ which satisfies both the configuration constancy constraint (i.e., remains unchanged) and $P_1$. To determine this, we check the satisfiability of
    $\unroll(\mathcal{S},2) \wedge \config(\cfg@0,\dots,\cfg@2)
    \wedge P_1(a@0,a@1, x@0,\dots,x@2)$,
    which expands to:
    \begin{align*}
    &x@0 = 0 \ \wedge \\
    &x@1 = \ite(\cfg@0,x@0+a@0,x@0-a@0) \ \wedge\\
    &x@2 = \ite(\cfg@1,\allowbreak x@1+a@1,x@1-a@1) \ \wedge\\
    &\cfg@1 =\cfg@0 \wedge \cfg@2=\cfg@1 \ \wedge\\
    &a@0\! =\! 1 \wedge a@1\! =\! 1 \wedge x@0\! =\! 0 \wedge x@1\! =\! 1 \wedge x@2\! =\! 2
    \end{align*}
    \noindent
    The formula is satisfiable when $\cfg@0 =\true$.
    
    \item $P_2(a@0, a@1, x@0, x@1, x@2) = a@0 = 1 \wedge a@1 = 1 \wedge x@0 = 0 \wedge x@1 = 1 \wedge x@2 = 0$. For this case, the formula to be checked is:
    \begin{align*}
    &x@0 = 0 \ \wedge\\
    &x@1 = \ite(\cfg@0,x@0+a@0,x@0-a@0) \ \wedge\\
    &x@2 = \ite(\cfg@1,x@1+a@1,x@1-a@1) \ \wedge\\
    &\cfg@1=\cfg@0 \wedge \cfg@2=\cfg@1 \ \wedge\\
    &a@0 = 1 \wedge a@1 = 1 \wedge x@0 = 0 \wedge x@1 = 1 \wedge x@2 = 0
    \end{align*}
    This formula is unsatisfiable, and thus there is no value of $\cfg$ that satisfies the desired property.
\end{enumerate}

\end{example}

\begin{figure}[t]
    \centering
    \includegraphics[scale=.4]{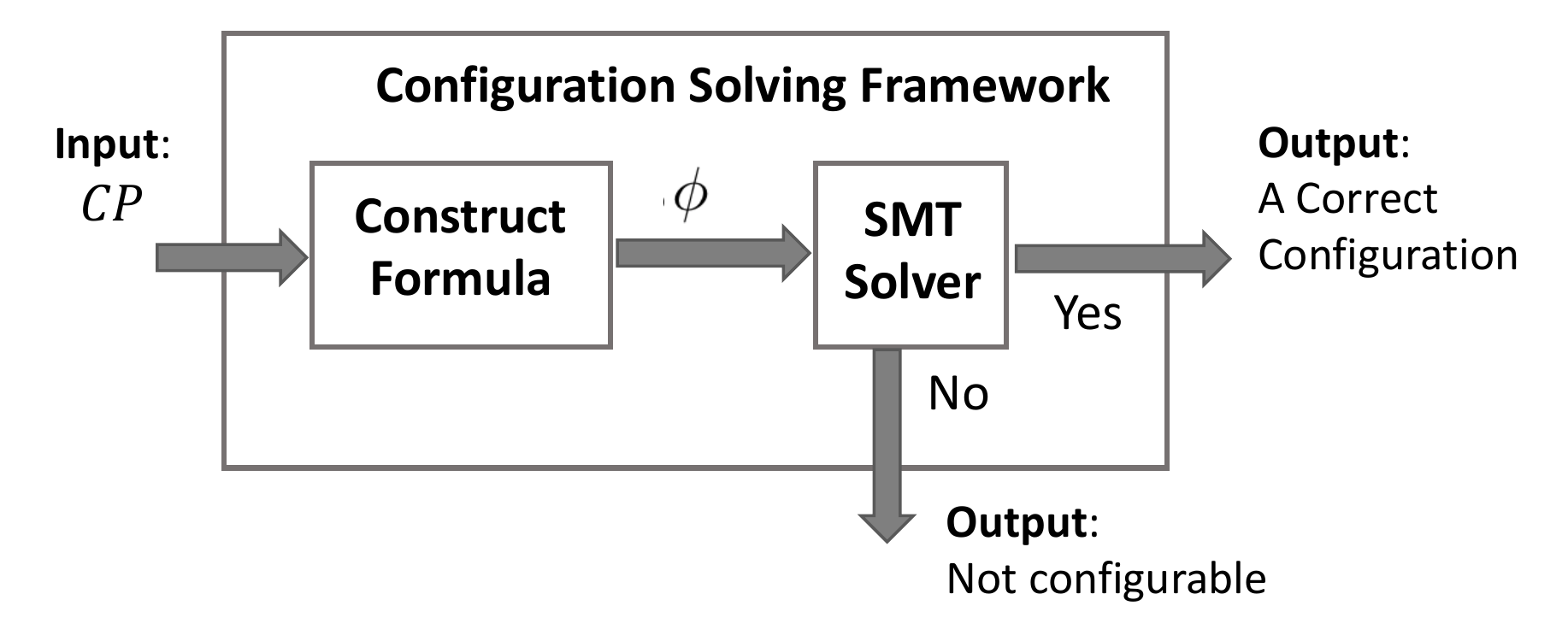}
    \caption{\small{Configuration solving framework (basic) scheme. $\mathcal{CP}$ is a configuration problem. $\phi$ is a configuration formula.}}
    \label{fig:frw}
\end{figure}

The framework for the basic scheme just outlined is shown in Figure~\ref{fig:frw}.  The input to the framework is a configuration problem.
The framework constructs formula~\eqref{F} and calls a solver to determine whether it is satisfiable.  The output is either ``not configurable'' or the configuration $\mathcal{C}$.

There are two main sources of complexity that limit the scalability of the approach.  The first is the complexity of the design itself, and the second is the bound $k$ required by $P$.  To address design complexity, we propose designing for modular configuration, discussed in more detail in Section~\ref{sub:botleneck_system} below.  Designing systems that can be configured using only small values of $k$ is an interesting research challenge that we plan to investigate in future work.

Another way to improve scalability is by using design knowledge to strengthen the formula $\phi$.  For example, if a configuration variable must be within a specific range, then this can be added as a constraint.  Any constraint expressible in the language supported by the backend SMT solver can be supported.

\subsection{Modular Configuration}
\label{sub:botleneck_system}

A natural remedy for design complexity is modular decomposition.
Here, we explain a systematic approach for modular configuration, including conditions under which a full configuration can be recovered.

Given $\mathcal{CP} = \langle \mathcal{S}\!, k,\! V_{\mathsf{in}},\! V_{\mathsf{out}},\! V_{\mathsf{conf}},\! P \rangle$ with $\mathcal{S}= \langle V, I, T \rangle$, we say $(\mathcal{CP}_1,\mathcal{CP}_2)$ is a \emph{decomposition} of $\mathcal{CP}$ (where
$\mathcal{CP}_i:=\langle \mathcal{S}_i, k, V^i_{\mathsf{in}}, V^i_{\mathsf{out}}, V^i_{\mathsf{conf}}, P_i \rangle$ and
$\mathcal{S}_i:= \langle V_i, I_i, T_i \rangle$ for $i=1,2$)
if: (i) $T_1(V_1,V'_1)\wedge T_2(V_2,V'_2) \implies T(V,V')$; (ii) $I_1(V_1) \wedge I_2(V_2) \implies I(V)$; (iii) $P_1 \wedge P_2 \implies P$; and  (iv) $V_{\mathsf{conf}} \subseteq V^1_{\mathsf{conf}} \cup V^2_{\mathsf{conf}}$.
%\nt{fix the formatting}

\begin{algorithm}[t]
%\algsetup{linenosize=\tiny}
\footnotesize 
%\hrule
\begin{algorithmic}[1]
 \procedure{\textsc{SolveModular}}{}
    \Input{
    $(\mathcal{CP}_1,\mathcal{CP}_2)$ a decomposition of $\mathcal{CP}$.
    }
   \Output {a pair $(r, \mathcal{C})$ where if $r = sat$, then $\mathcal{C}$ is a configuration of $\mathcal{S}$}
   \State $\phi_1 := \textsc{MakeCP}(\mathcal{CP}_1)$
   \State $(r, \Model_1) :=  \textsc{Solve}(\phi_1)$,\label{solve1}
    \If{$r = sat$}
    \State $\phi_2 := \textsc{MakeCP}(\mathcal{CP}_2)\wedge\textsc{GetAbduct}(\phi_1,\Model_1)$
    \label{propagate}
    \State $(r,\Model) := \textsc{Solve}(\phi_2)$\label{alg:solve_phi2}
    \EndIf
    \State \textbf{return} $(r,\Model^{V_\mathsf{conf}})$%\label{extract}
\end{algorithmic}
%\hrule
\caption{Modular configuration finding.}
\label{alg:modular}
\end{algorithm}

We now describe a procedure \textsc{SolveModular}, presented in Algorithm~\ref{alg:modular}, which, given a decomposition $(\mathcal{CP}_1,\mathcal{CP}_2)$ of a configuration problem $\mathcal{CP}$, attempts to solve $\mathcal{CP}$ by solving $\mathcal{CP}_1$ and $\mathcal{CP}_2$.
The call to $\textsc{MakeCP}$ on line 1 constructs the configuration formula for 
$\mathcal{CP}_1$.
The call to \textsc{Solve} on line 2 invokes a solver to check the satisfiability of the configuration formula. If the formula is satisfiable, \textsc{Solve} returns a pair $(sat,\Model)$ where $\Model$ is a satisfying interpretation found by the solver. If the formula is unsatisfiable, \textsc{Solve} returns a pair $(unsat,\Model)$ where $\Model$ is an arbitrary interpretation. Line 4 creates the configuration formula for $\mathcal{CP}_2$.  The formula is additionally constrained to ensure that the solution for $\mathcal{CP}_2$ still satisfies $\phi_1$.  The call to
$\textsc{GetAbduct}$ returns a formula $\psi$ such that $\psi \models_{\T} \phi_1$.  The goal is to use the information in $\Model_1$ to generate a simple formula for $\psi$.  The approach we take is to find a set of sub-terms in $\phi_1$ such that, if we constrain them to be equal to their values in $\Model_1$, this ensures that $\phi_1$ is satisfied.  In the worst case, we could constrain $\phi_1$ itself to be equal to $\top$, which would effectively require solving all of $\phi_1$ again at the same time as solving $\phi_2$.  However, in practice, we can do much better.  For example, it is often sufficient to let $\psi$ be the formula that assigns the free variables in $\phi_1$ to their model values from $\Model_1$.\footnote{See the appendix for details on when and why this works. Investigating other possible implementations for $\textsc{GetAbduct}$ is an interesting direction for future work.}
If the second call to \textsc{Solve} succeeds, the result is a correct configuration for $\mathcal{CP}$.

\begin{theorem}(Soundness)\newline
If $(\mathcal{CP}_1,\mathcal{CP}_2)$ is a decom\-position of a configuration problem $\mathcal{CP}$, and $\textsc{SolveModular}(\mathcal{CP}_1,\mathcal{CP}_2)$ returns a a pair $(sat,\mathcal{C})$, then $\mathcal{C}$ is a correct configuration of $\mathcal{CP}$.\label{Theorem}
\end{theorem}

%\IEEEproof
\begin{proof}
Let \textsc{SolveModular} return $(sat,\Model^{V_\mathsf{conf}})$. We prove that $\Model^{V_\mathsf{conf}}$ is a correct configuration of $\mathcal{CP}$. First, we notice that \textsc{SolveModular} returns $r = sat$ iff both calls to $\textsc{Solve}(\phi_1)$ and $\textsc{Solve}(\phi_2)$ return $r = sat$. Let $(sat,\Model_1)$ and $(sat,\Model)$ be the results of $\textsc{Solve}(\phi_1)$ and $\textsc{Solve}(\phi_2)$, respectively. Let $\psi = \textsc{GetAbduct}(\phi_1,\Model_1)$. From line~\ref{alg:solve_phi2}, $\Model \models \phi_2$. Thus, $\Model \models \textsc{MakeCP}(\mathcal{CP}_2)$ and $\Model \models \psi$. Since $\psi \models_\mathcal{T} \phi_1$, we also have $\Model \models \phi_1$.  Consequently, $\Model$ satisfies: $I_1$, $T_1(V_1@i,V_1@(i+1))$ for $i\in[0,k-1]$, $\config(V^1_{\mathsf{conf}},k)$, and $P_1$. Furthermore, $\Model$ satisfies: $I_2$, $T_2(V_2@i,V_2@(i+1))$ for $i\in [0,k-1]$, $\config(V^2_{\mathsf{conf}},k)$, and $P_2$. 
By the definition of decomposition, then, $\Model$ satisfies $I(V)$, $T(V@i,V@(i+1))$ for $i\in[0,k-1]$, and $P$. 
Finally, from $\Model \models \config(V^1_{\mathsf{conf}},k)$, $\Model \models \config(V^2_{\mathsf{conf}},k)$, and condition (iv) of the definition of decomposition ($V_{\mathsf{conf}} \subseteq V^1_{\mathsf{conf}} \cup V^2_{\mathsf{conf}}$), it follows that $\Model \models \config(V_{\mathsf{conf}},k)$. Thus,  $\Model$ satisfies the configuration formula of $\mathcal{CP}$. Therefore, $\mathcal{C}:=\Model^{V_{\mathsf{conf}}}$ is a correct configuration of $\mathcal{CP}$.
\end{proof}

If \textsc{SolveModular}  returns $r = unsat$, this does not (in general) imply that $\mathcal{CP}$ is unconfigurable. Rather, it may be that the particular decomposition fails, or even that the particular solution found for $\mathcal{CP}_1$ is at fault (and another solution would have succeeded).
%\nt{shall we reword the last sentence? Or drop?}
%\IEEEQEDopen

\begin{figure}[t]
  \centering
  \includegraphics[scale=.45]{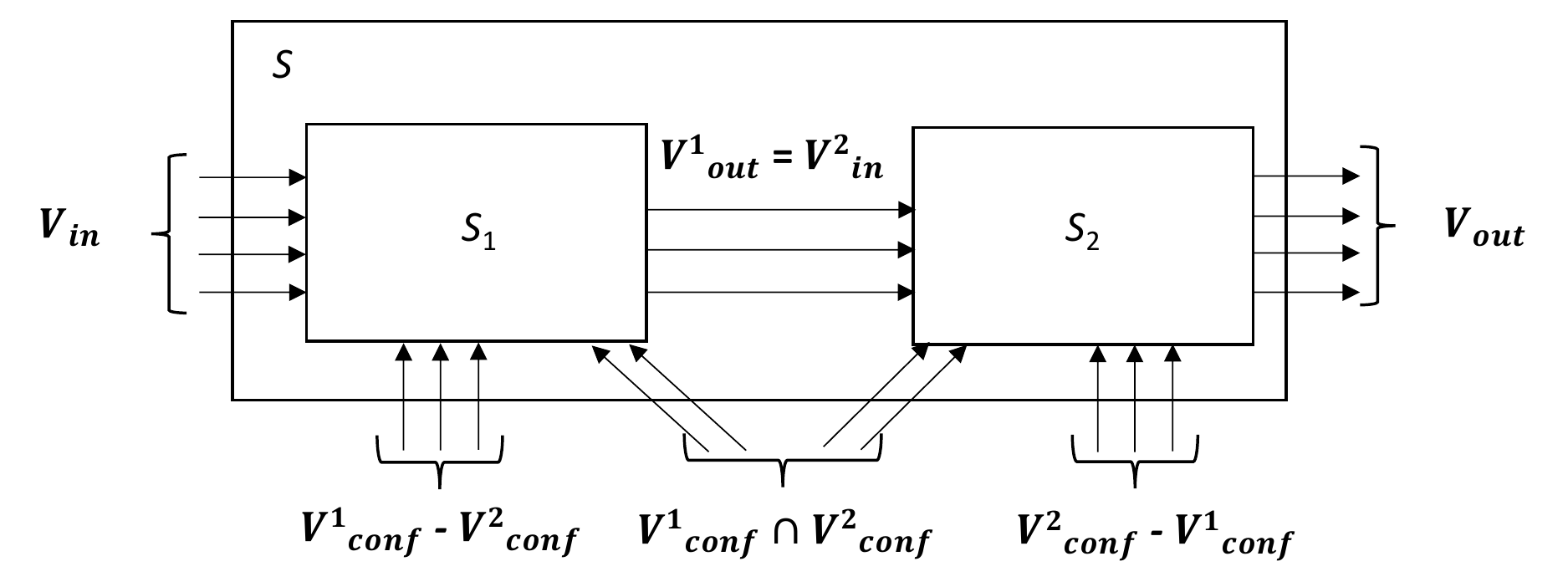}
  \caption{\small{Modular decomposition of system $\mathcal{S}$ into systems $\mathcal{S}_1$ and $\mathcal{S}_2$. $V^1_{out}$ and $V^1_{conf}$ are the output and the configuration variables of $\mathcal{S}_1$. $V^2_{in}$ and $V^2_{conf}$ are the input and the configuration variables of $\mathcal{S}_2$.  $ V_{conf}\subseteq V^1_{conf} \cup V^2_{conf}$.}}
  \label{fig:decomposed}
\end{figure}

However, in practice, we have found that the algorithm works well when the decomposition separates a module into two largely independent parts.  An example is shown in Figure~\ref{fig:decomposed}.  Here, the two submodules share only a subset of the configuration variables as well as an interface where outputs of the first module flow into inputs of the second module.

\section{Optimization-Assisted Configuration}
\label{section:optimization}

\begin{figure*}[ht]
    \centering
    \includegraphics[width=4\textwidth/5]{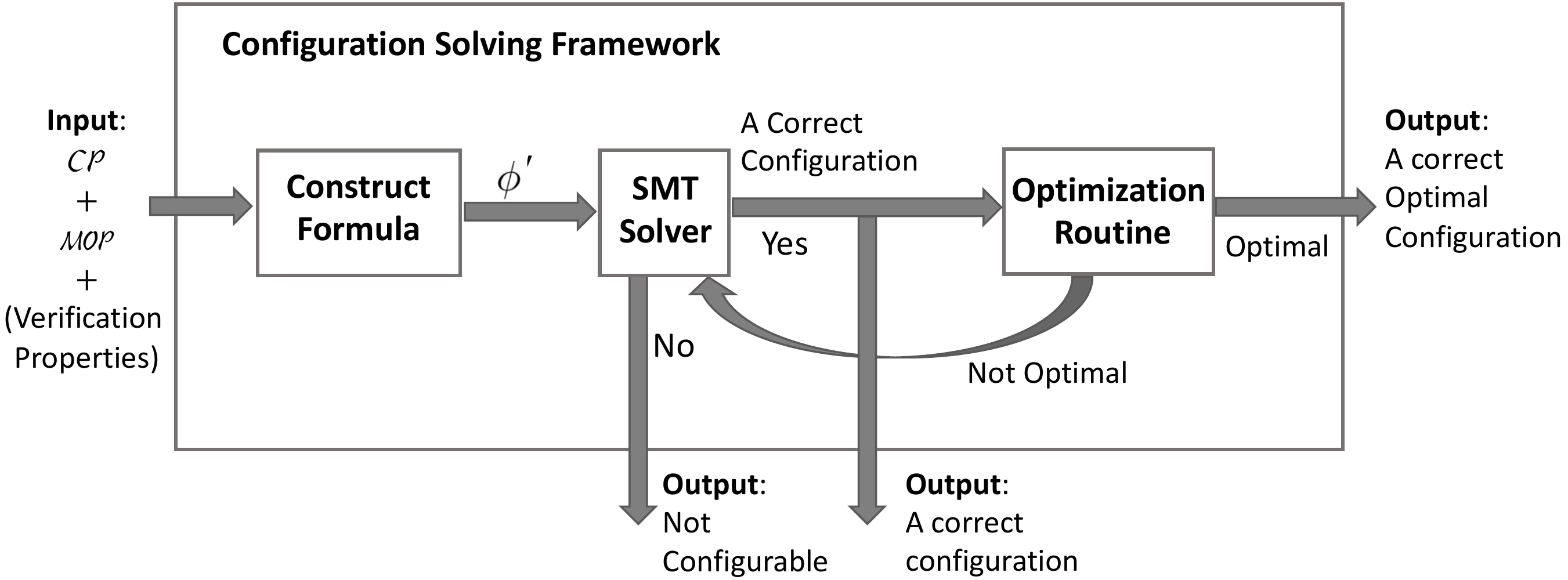}
    \caption{\small{Optimization-assisted configuration framework. The input is a configuration problem with optional optimization and verification objectives. The framework can return: (i) a non-optimal but correct configuration, or (ii) an optimal and correct configuration, or (iii) $\textsf{unsat}$. $\phi'$ is a conjunction of the configuration formula $\phi$ and the optional verification properties.}}
    \label{fig:frwstable}
\end{figure*}
A solver can return an unnatural or non-intuitive configuration, complicating the ability of users to understand or maintain the configuration.

We observe that users tend to prefer the simplest configurations, where the notion of simplest corresponds to minimizing some metric when finding solutions. 
To this end, we show how to extend our framework with optimization goals.

Figure~\ref{fig:frwstable} depicts our configuration framework extended with support for multi-objective optimization. There are various ways to combine an optimization routine with the configuration solving, we depict one such approach. The optimization routine implements an iterative optimization approach, such as a branch-and-bound algorithm. In the branch-and-bound algorithm first a solution is found and the value of the objective term is calculated; then the search space is systematically explored by iteratively constraining the value to be better than the current best value. There are many different kinds of optimizations that fit this general framework.  We will present several useful examples in the context of the case study in Section~\ref{section:CaseStudy}.

\vspace{1ex}
\noindent
{\bf Further extensions.}
Figure~\ref{fig:frwstable} also includes an extension to support the verification of additional properties that the system should have. In this scheme, simple invariant properties are conjoined to the configuration formula and it is ensured that
the configuration found satisfies the invariant up to bound k. To check that an invariant holds for all time, would require a separate unbounded model check after configuration. This approach can be implemented as a feedback loop alternating the configuration finding and verification queries. 

Lifting our configuration finding technique to an unbounded setting is an interesting direction for future work. We will briefly discuss some challenges and ideas for using existing unbounded model-checking approaches, such as invariant synthesis. First of all, it would require writing the configuration property as an invariant which may be much harder than writing it as a set of input, output pairs. If this is possible, then we can utilize invariant synthesis techniques by querying to synthesize an invariant in this form: $\bigwedge_i V_{conf}^i = C^i \implies P$, where the left-hand-side of the implication contains all configuration variables $V_{conf}^i \in V{conf}$ and $C^i$ is a constant value to be synthesized, $P$ is the input-output property for the unbounded case. Note, that this formula is indeed an invariant if $C^i$ values are a correct configuration for the property $P$.

\section{Case Study}
\label{section:CaseStudy}

We present a case study with a course-grained reconfigurable architecture (CGRA) design developed in the Agile Hardware Center at Stanford University~\cite{AHA}. Reconfigurable architectures are appealing because they offer the high performance of hardware with software-like flexibility.  CGRAs in particular use sophisticated reconfigurable elements with the aim of narrowing the performance gap with custom ASICs~\cite{10.1145/3357375}.

However, configuring a CGRA is challenging, typically requiring manual effort by an experienced engineer who fully understands the application and the design. To the best of our knowledge, ours is the first framework that finds correct CGRA configurations fully automatically.

In this paper, we focus on configuring a \emph{memory tile} of the CGRA for image processing applications.  In these applications data is streamed into the memory tile and must be reordered in various ways before being streamed out.  Only the timing and order of the data are changed; the data itself remains the same.  Below, we first describe the memory tile design, then present some specific applications, and then explain how we automate configuration of the design for these applications.

\subsection{CGRA Memory Tile Design}

\begin{figure*}[ht]
    \centering
    \includegraphics[width=4\textwidth/5]{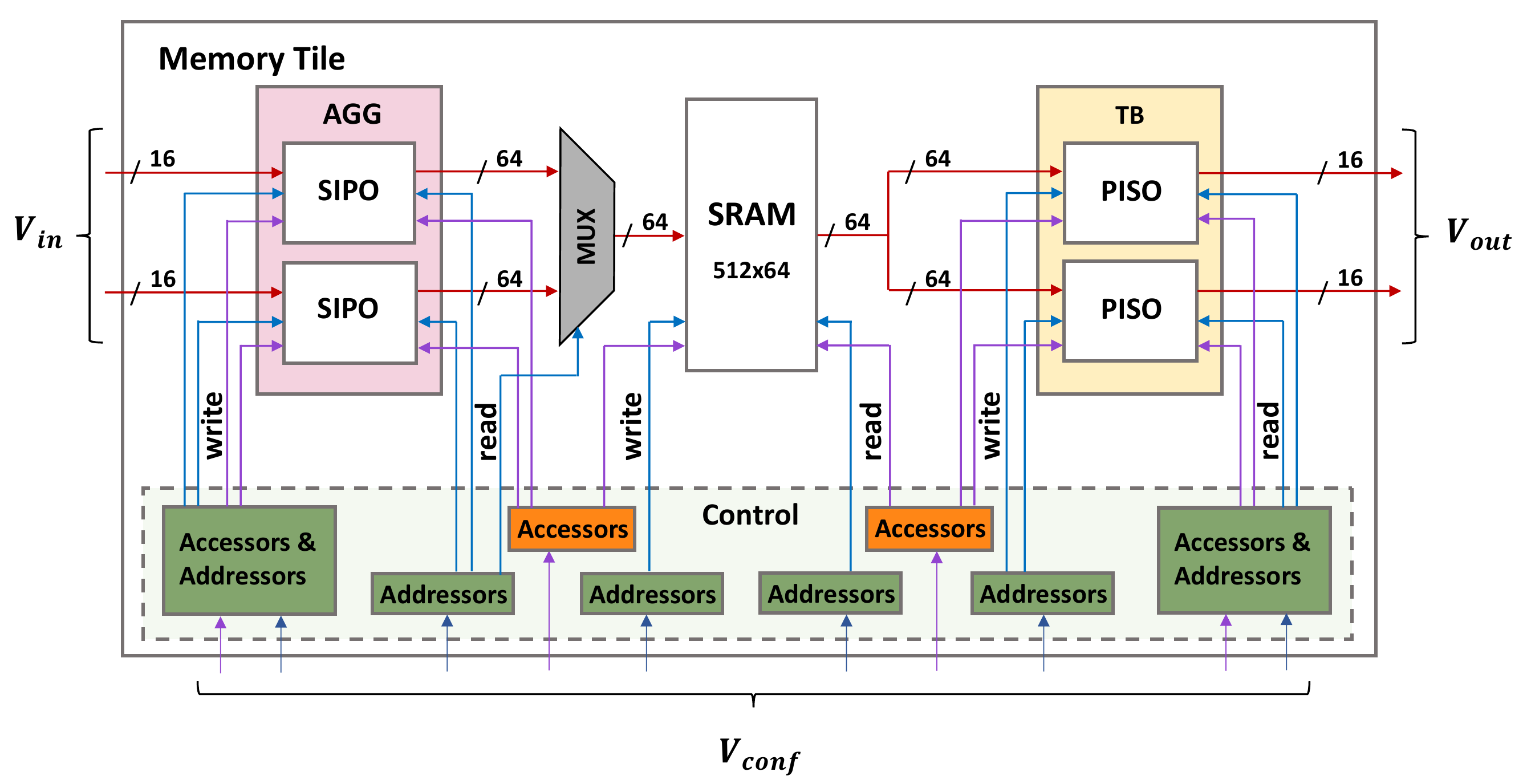}
    \caption{\small{Memory tile architecture. All accessors and addressors are included in the \emph{control} box. Red arrows represent data flow. Blue and purple arrows represent addressor and accessor control signals, respectively. 
    Green boxes are local to a single module. Orange boxes are shared between modules. 
    $V_\mathsf{conf}$ consists of all accessor and addressor configuration variables.}}
    \label{fig:tile1}
\end{figure*}

The memory tile is a non-trivial design (34998 FF and 164696 gates).  Figure~\ref{fig:tile1} shows its architecture . It contains three types of units: \emph{memories}, \emph{addressors}, and \emph{accessors}. Addressors and accessors are reconfigurable units. The accessors control \emph{when} to write or read. 
The addressors control \emph{where} to write or read. 
There are three memory modules: an \emph{aggregator} module (AGG), a \emph{static random-access memory} module (SRAM), and a \emph{transpose buffer} module (TB). Each module has an \emph{input accessor} and an \emph{input addressor} associated with it for writes, and an \emph{output accessor} and an \emph{output addressor} for reads. The modules are chained: outputs of AGG are intputs to SRAM, and outputs of SRAM are inputs to TB. Accessors are \emph{shared} between each pair of connected memory modules.  Shared accessors act as \emph{schedule generators} for each memory connection. They specify when the data should be transferred and set any required delays between when the data is produced and consumed.  Addressors are unique for each module.

The addressors and accessors in the memory tile make use of affine sequence generators to generate sequences of values for reading and writing.  Figure~\ref{fig:mekloops} shows pseudocode for an affine sequence generator.  It takes as input a number $\dimensionality$ of loops, an array $\ranges$ with bounds for each loop, an array $\strides$ with strides for each loop, and $\startingaddr$ which is a base value.  It then computes a sequence of outputs, $\vals$, by running $\dimensionality$ nested loops, and computing the sum of the offset and the product of each stride with its loop index in the innermost loop.
Each of the inputs to the procedure corresponds to a configuration register in the hardware.

\begin{figure}[t]
\hrule \vspace{2pt}
\begin{algorithmic}[1]
\footnotesize 
 \procedure{\textsc{AffineSequence}}{}
    \Input{
    $\dimensionality$: a value indicating the number of nested loops,\newline
    $\ranges[\dimensionality]$: an array of loop bounds, one for each loop,\newline
    $\strides[\dimensionality]$: an array of strides, one for each loop,\newline
    $\startingaddr$: the offset for the address computation
    }
   \Output {$\vals[\Pi_i \ranges[i]]$: a set of output addresses}

\State{var $c[\dimensionality];$} \Comment{Index variables for each loop} 
\State{var $i := 0;$}
\For{$c[\dimensionality-1]  \textbf{ in }[0, \ranges[\dimensionality-1])$}
\State{$...$}
\For{$c[0] \textbf{ in }[0, \ranges[0])$}
\State{$\vals[i] := \Pi^{\dimensionality-1}_{j=0}\vect{c}[j] * \strides[j]+\startingaddr;$}
\State{$i := i + 1;$}
\EndFor
\EndFor
\end{algorithmic}
\hrule
\caption{\small{Affine sequence generator using nested loops.}} 
\label{fig:mekloops}
\end{figure}

While each addressor and accessor contains an affine sequence generator, they differ in how they interpret $\vals$. 
For an addressor, $\vals$ contains raw addresses sent to a memory (for either reading or writing). For an accessor, $\vals$ contains clock cycle counts that are compared to a running cycle counter to determine when to read or write. 
Note that an (accessor, addressor) pair should have the same values for their $\dimensionality$ and $\ranges$ variables to ensure that they produce the same number of values.
There are 4 accessors (including 2 shared with SRAM) and 4 addressors for AGG (1 for each memory port). TB has 4 accessors (including 2 shared with SRAM) and 4 addressors (1 for each memory port). SRAM has 2 addressors, and shares 2 accessors with AGG and 2 acessors with TB.

The memory tile processes 16-bit words. However, it uses a 512x64-bit SRAM which stores four 16-bit words at each address.  The rationale for this design is to emulate a multi-ported SRAM while minimizing the energy consumption per memory access~\cite{vasilyev2019evaluating}. To match the data width at the SRAM interface, AGG and TB implement width converters. AGG implements a \emph{serial-in to parallel-out} (SIPO) converter---serial data is loaded, one 16-bit word at a time, and these are packed into 64-bit outputs. TB implements a \emph{parallel-in to serial-out} (PISO) converter---parallel data is loaded into the PISO as a 64-bit word and is shifted out of the PISO serially, one 16-bit word at a time. 
The memory tile uses a 2-input and 2-output port architecture to support more throughput. Thus, AGG and TB contain two SIPOs and two PISOs, respectively.

\subsection{Stencil Applications}
We consider a common class of image-processing techniques called \emph{stencils}. Stencil computations usually consist of a multi-stage pipeline, where each stage is a dense linear algebra computation in a local region. So-called \emph{push memories} are inserted between computation units, whose job is to orchestrate the order and the timing of the data explicitly~\cite{buffets}.  We explore configuring memory tiles as push memories for four stencil applications:

\begin{itemize}
\item  \emph{Identity.}  The identity stencil simply streams the input back out in the same order.  It is useful as a baseline test and also can be used to implement a fixed delay on a stream.

\item \emph{3x3 Convolution.} This stencil is used in a variety of image processing applications~\cite{chandel2013image} (e.g., to blur images).  It multiplies a 3x3 sliding image window by a 3x3 kernel of constant values.
\item \emph{Cascade.}  This application implements a pipeline with two convolution kernels executed in sequence. 
The Cascade application requires configuration of two memory tiles, denoted by \emph{conv} and \emph{hw}.
\item \emph{Harris.}  Harris is a corner detection algorithm that can be used to infer image features~\cite{Harris1988combined}. It extracts the gradients of an image in different orientations and combines this information using multiple convolutions. This is the most complex of our applications, requiring the configuration of five different memory tiles, which we denote as
 \emph{cim}, \emph{lxx}, \emph{lxy}, \emph{lyy}, and \emph{pad}.
\end{itemize}

\subsection{Automating the Memory Tile Configuration}
\label{subsection:human-understandable-configs}

We decompose the memory tile into three sub-modules (for scalability), following the approach shown in Figure~\ref{fig:decomposed}. The first sub-module includes AGG, its input/output accessor/addressor modules, and the MUX (1372 FF, 19676 gates). The second sub-module includes SRAM, both AGG read accessors, and both TB write accessors (33712 FF, 150750 gates). The third sub-module includes TB and its input/output accessor/addressor modules (1126 FF, 18538 gates). Shared accessors contain the shared configuration variables, whose values are propagated to the next module during modular configuration.

In order to configure each module in the memory tile, we look at the transition system defined by its memory and its accessors and addressors.  We then implement a ``programming by example'' approach to specify the input-output property $P$: we use a sequence of distinct input values (e.g., 1,2,3,\dots), paired with the corresponding application-specific desired output sequence of those same values to define the property, and solve for the configuration variables as described in Section~\ref{subsec:cp} above. This allows us to avoid universally quantifying the input and output variables in $P$, as required in the classic formulation of the configuration finding problem.

As mentioned in Section~\ref{section:optimization}, it is important to generate configurations that can easily be read and understood.  Working together with the designers, we devised a set of optimization objectives that greatly improve the readability of memory tile configurations.  We explain these next. We apply the framework of Figure~\ref{fig:frwstable} to configure and optimize each module separately.

\emph{Objective} 1: we first minimize the $\dimensionality$ variables in the module, since this corresponds to using fewer nested loops and fewer loop counters, resulting in simpler solutions in general. 
We prioritize minimizing $\dimensionality$ variables controlling writes over those controlling reads, as lower write complexity leads to lower read complexity anyway. 
We formalize this as the following multi-objective optimization problem:\\ 
\resizebox{\linewidth}{!}{
  \begin{minipage}{\linewidth}
\begin{align*}
\mathcal{MOP}_1 &:= \{\mathcal{OP}_1,\mathcal{OP}_w^{1}, \dots,\mathcal{OP}_w^{d_w}, \mathcal{OP}_r^{1}, \dots,\mathcal{OP}_r^{d_r}\}:\\
\mathcal{OP}_1&:= \langle  \Sigma_i \; \dimensionality_i,A_{BV},\preccurlyeq_{BV}, \phi, \mathit{min} \rangle \text{ for }i \in [1,d],\\
\mathcal{OP}_w^i&:= \langle \dimensionality_w^i,A_{BV},\preccurlyeq_{BV}, \phi, \mathit{min} \rangle \text{ for }i \in [1,d_w]\\
\mathcal{OP}_r^i&:= \langle \dimensionality_r^i,A_{BV},\preccurlyeq_{BV}, \phi, \mathit{min} \rangle \text{ for }i \in [1,d_r]\\
\end{align*}
\end{minipage}
}

\noindent Here, $A_{BV}$ is the domain of bit-vectors (i.e., unsigned machine integers), $\preccurlyeq_{BV}$ is the usual total order on bit-vector values, 
$d$ is the number of affine sequence generators in the module, and $\dimensionality_i$ for $i\in[1,d]$ are all of the $\dimensionality$ variables in the module.  These are further partitioned into write dimensionality variables $\dimensionality_w^i$, $i\in[1,d_w]$,
and read dimensionality variables,
$\dimensionality_r^i$, $i\in[1,d_r]$, with $d_w+d_r=d$. 
$\phi$ is the configuration formula.

\emph{Objective} 2: we minimize the products of the range configuration variables in each loop-nest structure.  The objective term corresponds to the aggregate number of reads or writes that occur to a particular memory. By minimizing this number, we eliminate unnecessary reads and writes to the memory. 
Formally, the optimization problem is:
\[{\mathcal{OP}_{2}:=\langle   \Sigma_{i=0}^{d-1}\Pi^{\dimensionality_i-1}_{j=0}\ranges_i[j],A_{BV},\preccurlyeq_{BV}, \phi, \mathit{min} \rangle}\] 

\emph{Objective} 3: we minimize stride variables to avoid generating configurations using unnecessarily large addresses.
 
Many different sets of values for strides could produce the same $\vals$ stream in the end, so by choosing the smallest values, we hope to generate the simplest solution. The optimization problem simply minimizes the sum of all stride variables in the module: \[\mathcal{OP}_3:= \langle \Sigma_i \; \mathit{strides_i},A_{BV}\preccurlyeq_{BV}, \phi, \mathit{min} \rangle.\]

\emph{Objective} 4: we also minimize $\startingaddr$ configuration variables in addressor modules. For addressor modules, minimizing the $\startingaddr$ addressor variable prevents unnecessary offsets, improving the readability of the generated configuration. Note that values of $\startingaddr$ variables in the accessors are fixed by the application. The corresponding problem is as follows, minimizing the sum of all addressor $\startingaddr$ variables in the module:
\[\mathcal{OP}_4:= \langle \Sigma_i\;  \startingaddr_i,A_{BV},\preccurlyeq_{BV}, \phi, \mathit{min} \rangle.\]

\emph{Combined objective}: the combined optimization query includes all four objectives and captures the full set of optimization objectives for each module: 
\[\mathcal{MOP_H}:=\{\mathcal{MOP}_1,\mathcal{OP}_2,\mathcal{OP}_3,\mathcal{OP}_4\}.\]

We solve and prioritize $\mathcal{MOP}_1$ by iteratively increasing the bound on the sum $\Sigma_i \dimensionality_i$, and for each bound, trying all possible assignments to the variables, in the order specified by $\mathcal{MOP}_1$. Note, this approach does not directly fit in the scheme described in Figure~\ref{fig:frwstable}, since it does not require finding the first solution that is iteratively improved. Instead it iteratively widens the search space until the first solution is found.

For the other objectives, we use a branch-and-bound algorithm.
First, a solution is found, and the value of the term is calculated; then, the solution space is explored systematically, by iteratively constraining the value of the objective term to be better than the current best value. 
Each optimal solution is propagated to the next optimiziation objective as a constraint.

\section{Evaluation}
\label{section:evaluation}

\begin{figure*}[ht]
    \centering
    \begin{subfigure}{.45\textwidth}
        \centering
        \includegraphics[width=.38\paperwidth]{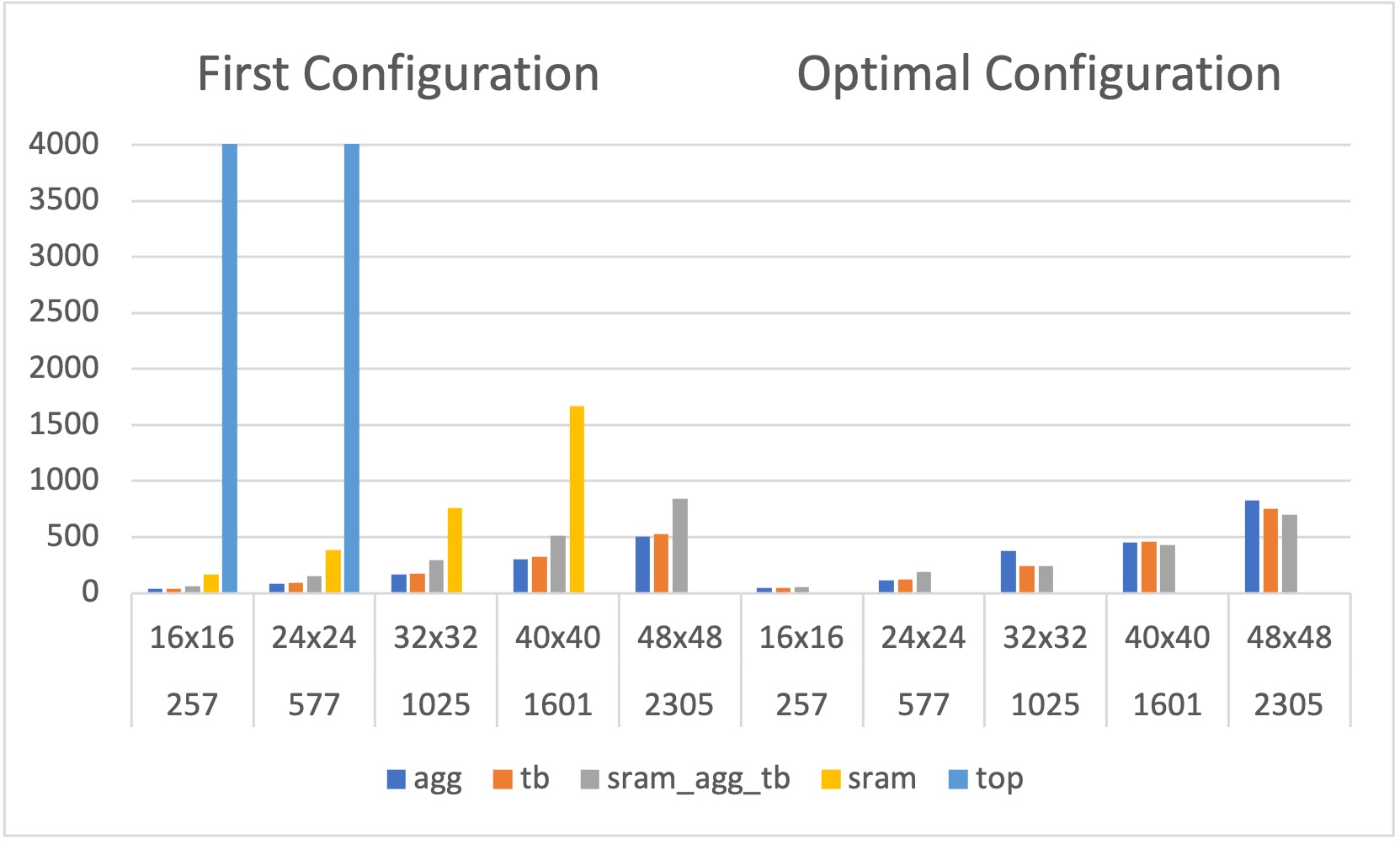}
        \caption{\small{Identity Stream}}
        \label{fig:isbaseline}
    \end{subfigure}
    \begin{subfigure}{.45\textwidth}
        \centering
        \includegraphics[width=.38\paperwidth]{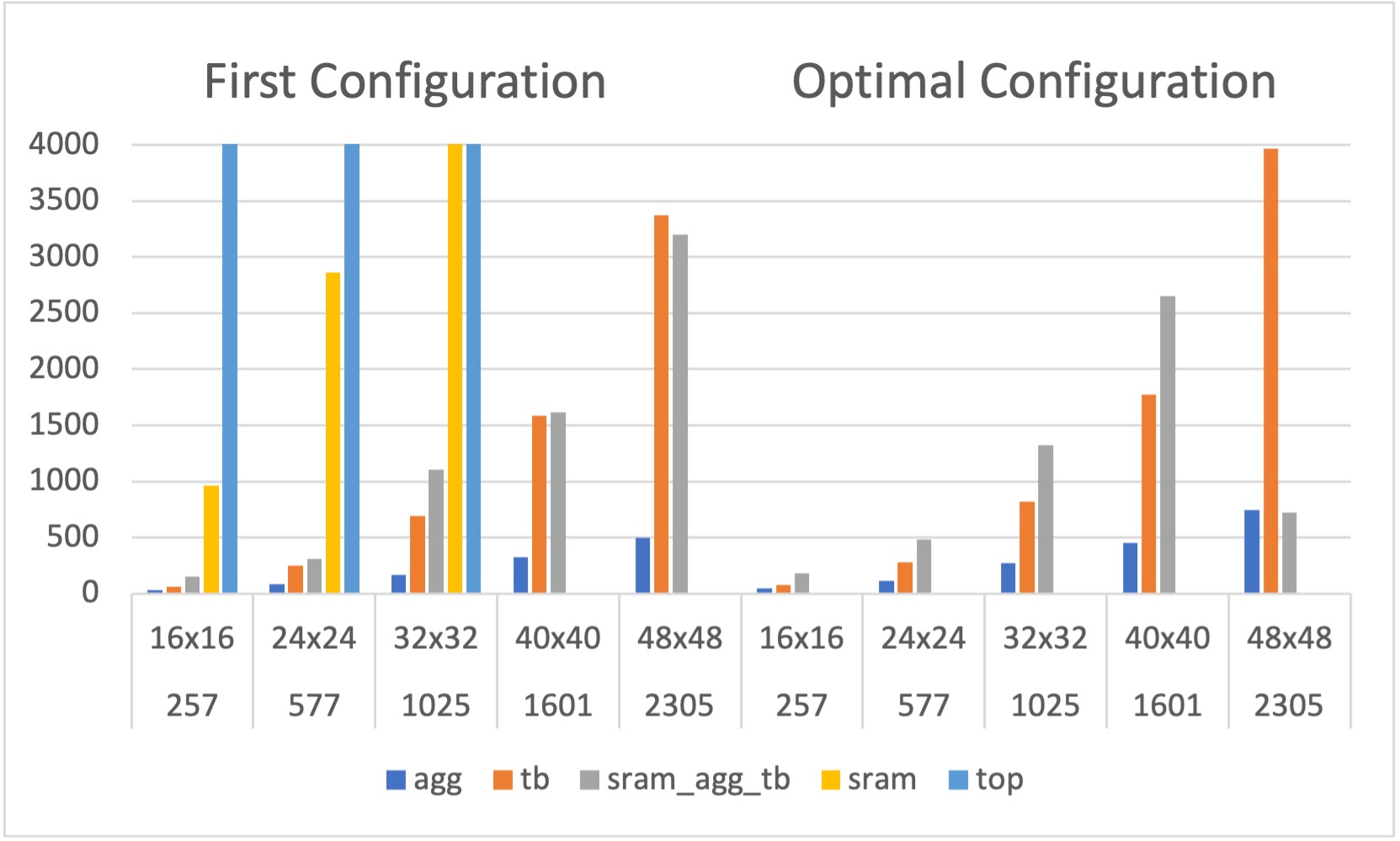}
        \caption{\small{3x3 Convolution}}
        \label{fig:isbest}
    \end{subfigure}
    \begin{subfigure}{.45\textwidth}
        \centering
        \includegraphics[width=.38\paperwidth]{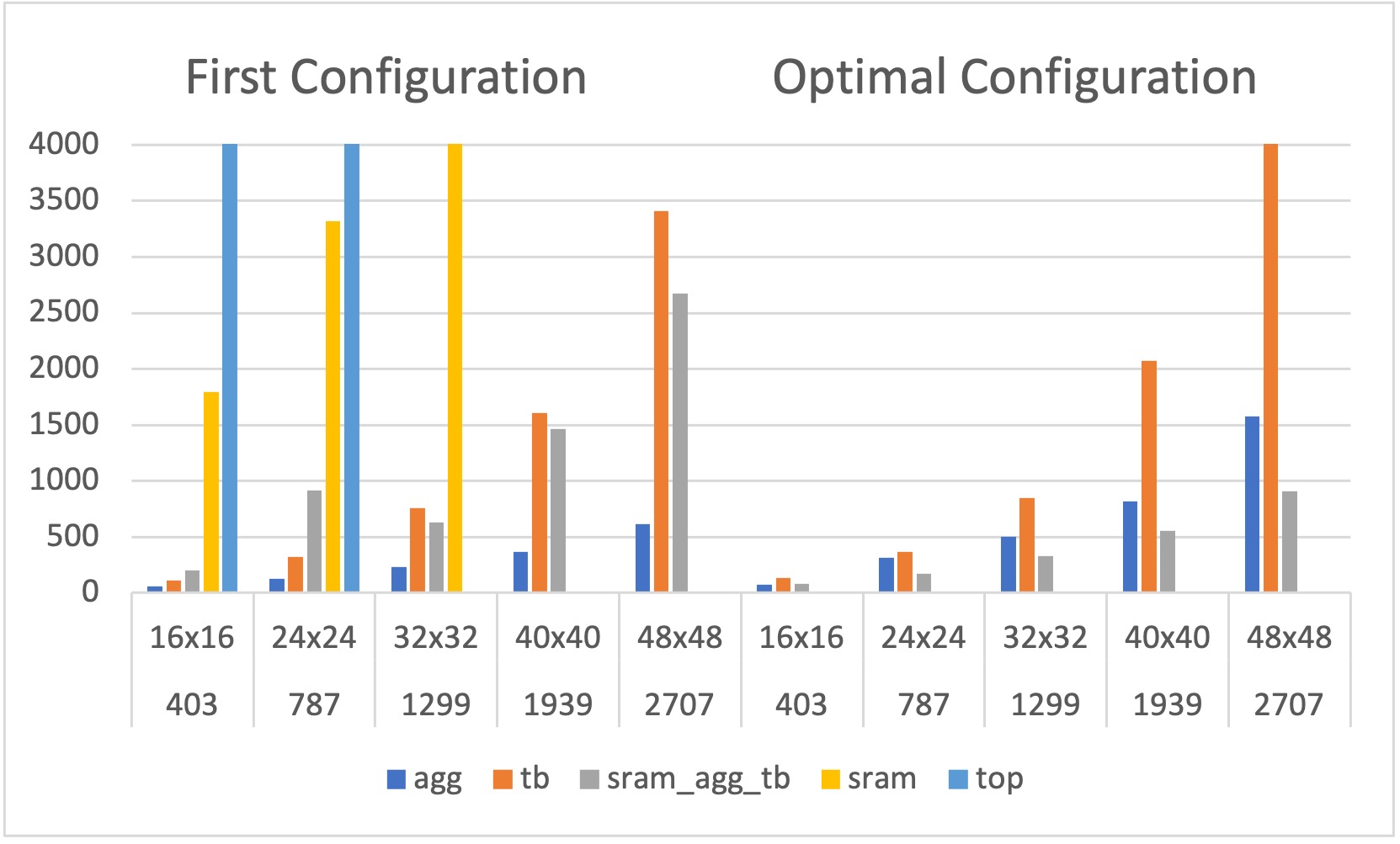}
        \caption{Cascade (conv)}
        \label{fig:Harrislxxbaseline}
    \end{subfigure}
    \begin{subfigure}{.45\textwidth}
        \centering
        \includegraphics[width=.38\paperwidth]{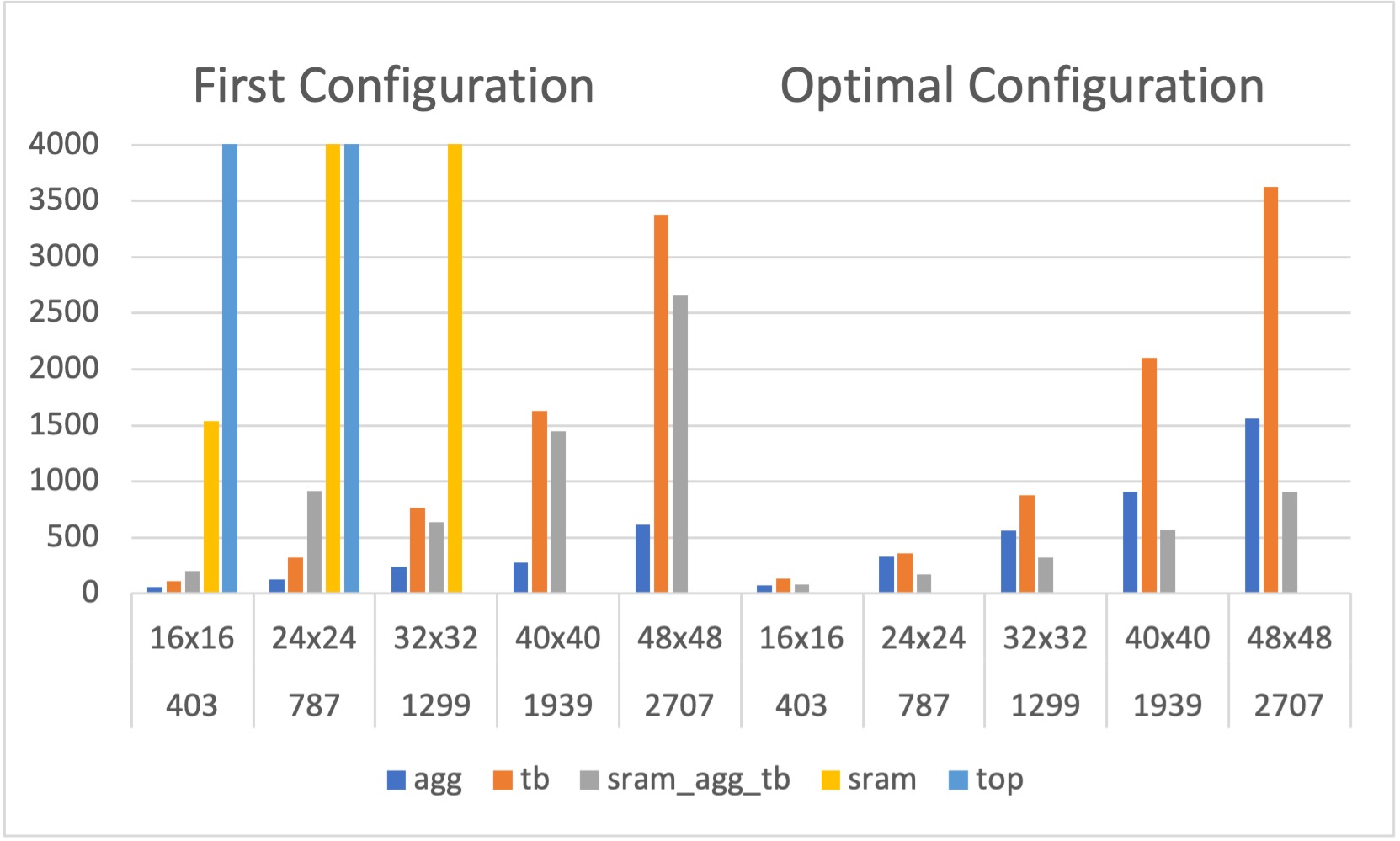}
        \caption{Harris (lxx)}
        \label{fig:Harrislxxbest}
    \end{subfigure}
    \caption{\small{Horizontal axis shows image sizes and number of clock cycles required for processing. Vertical axis shows time in seconds.}\vspace*{-1em}}
    \label{fig:baseline-best}
\end{figure*}

\vspace{1ex}
\noindent
{\bf Implementation.}
We have implemented our framework using \pono~\cite{pono}, an open-source SMT-based model checker. \pono is built on \smtswitch~\cite{smt-switch}, a generic C++ API for interacting with SMT solvers. \pono provides infrastructure for reading in, unrolling, and otherwise manipulating transition systems. 
We use \btor~\cite{DBLP:conf/cav/NiemetzPWB18} as the underlying SMT solver. We convert the memory tile design in our case study from a SystemVerilog representation to its equivalent representation in the Btor2 format~\cite{DBLP:conf/cav/NiemetzPWB18}, which is accepted by \pono. We use \yosys~\cite{yosys}, a Verilog synthesis suite, to do the translation.

\vspace{1ex}
\noindent
{\bf Experimental Results.}
We evaluate our configuration-finding framework using the memory tile design and the four stencil applications described in Section~\ref{section:CaseStudy}. 
For each application, we generated benchmarks for various input image sizes, from 16x16 to 60x60.  For applications that require more than one memory tile (i.e., cascade and harris), we choose one representative configuration problem: conv for Cascade and lxx for Harris (more results appear in the appendix).  The number of transitions required for each configuration problem is based on the number of clock cycles it takes to process an image of a given size for a given application.

For each benchmark, we first run the basic algorithm described in Section~\ref{section:framework}, which finds the first satisfying configuration.  We try both with and without the modular approach described in Section~\ref{sub:botleneck_system}.  We then run our optimization-assisted configuration algorithm (using only the modular approach) as described in Section~\ref{section:optimization}.
We run our experiments on a 2x Intel Xeon E5-2620 v4 @ 2.10GHz 8-core 128GB computer. Timeout is set to 4000 seconds. Memory limit is 100 GB.

The results are shown in Figure~\ref{fig:baseline-best}.  Each chart shows results for both the basic algorithm (First Configuration) and the optimization-assisted algorithm (Optimal Configuration).  Within each of these categories, up to five different results are shown for each image size: \emph{top} is the time required to configure the entire design, monolithically; \emph{agg}, \emph{tb}, and \emph{sram} refer to the time required to configure each of the sub-modules independently; and \emph{sram\_agg\_tb} is the time required to configure the SRAM module after first configuring AGG and TB (this is the most efficient order for these modules) and then propagating the shared configurations from those modules as described in Figure~\ref{fig:decomposed}.  Note, in the modular approach, AGG and TB are configured independently, thus the configuration can be performed in parallel and the total design configuration time is the sum of \emph{sram\_agg\_tb} and the maximum of \emph{agg} and \emph{tb}. Timeouts are represented by full bars (up to the timeout limit), and memory outs are represented by omitting the bar completely.  We also omit the bar for \emph{sram\_agg\_tb} if either AGG or TB is not solved within the given time-memory budget.  We make several observations about the results below.

{\bf Modular Approach}. As the experiments show, the full memory tile is too large to solve within the given time-memory budget---it times out for all image sizes.  
However, by using the modular approach, we are able to configure the design for all applications for reasonably useful image sizes. For the Identity Stream, we can configure for all image sizes (with unroll depths up to 3601) relatively easily using the modular approach. Other applications are more challenging, but we are still able to scale up to images of size 40x40 (and unroll depth up to 1939 clock cycles).

We also observe that the AGG and TB modules take comparable time for the Identity Stream, but for other applications, configuration of the TB module is more challenging.
This can be explained as follows. AGG and TB are both two-port designs, comparable in size and complexity. But for all applications, AGG can be configured by exploiting only a single port, while only the Identity Stream allows a single-port configuration of TB.  Thus, we quickly find a simple configuration for TB with the Identity Stream, but no comparatively simple configuration exists for the other applications.

\iffalse
\begin{figure}[t]
    \centering
    \includegraphics[width=1\linewidth]{All_optimized.jpg}
    \caption{\small{Human-understandable configuration for Identity Stream, 3x3 Convolution, Cascade, and Harris applications. Horizontal axis shows the number of clock cycles required to process a 60x60 image by each application.}}
    \label{fig:alloptimal}
\end{figure}
\fi

{\bf Optimal Configurations}. 
The right-hand side of each chart shows the results of running our optimization-assisted configuration algorithm for each application.  There are several interesting observations. First of all, for the AGG and TB modules, finding optimal configurations is generally more expensive.  However, once these optimal configurations are found, it is often easier to find the corresponding SRAM configuration, suggesting that optimal configurations may help improve later stages of modular configuration.  The total configuration time with optimization is generally comparable to or only slightly worse than the time required to configure without optimization.  Given the value of optimal configurations in terms of simplicity and  readability, these results suggest that modular configuration with optimization may be the best strategy in practice.

\iffalse
Figure~\ref{fig:alloptimal} illustrates finding human-understandable configurations for 60x60 pixels\footnote{For experimental results with smaller image sizes see the Appendix.}. In this setup, $\mathcal{MOP_H}$ is applied and the auxiliary design-specific constraints are added. These constraints are same as in the \emph{Auxiliary Constraints} heuristic. Every task, except from Harris (pad) and Harris (cim) has completed within the 3-hour timeout. Harris (pad) instance took a little over 3 hours for the image size 60x60. \nt{Harris (cim) timed-out for the image 46x46. not visible in Figure 8.} Optimal configuration of TB took the longest time in most of the applications. For the Identity Stream, all tasks were completed within 33 minutes.
\fi

\section{Related Work}
\label{section:related}

The problem of system configuration has been studied in various formulations and domains, such as software tool configuration, hardware configuration, network configuration, distributed application configuration, and deployment strategies. 
In one research stream, the configuration problem is to select and arrange a set of components from a given set of assets in order to construct an overall system with a desired specification~\cite{DBLP:journals/ai/McDermott82,Mansor2016VLSICC,DBLP:conf/cade/MichelHGH12,Sabin1996ConfigurationAC}. 
Other formulations take as input a configuration database, including configuration variables, and desired requirements to be met~\cite{DBLP:journals/jnsm/NarainLMK08,Narain2005NetworkCM}. The task is to find values for the configuration variables which instantiate the database so that it meets the requested requirement. The work whose problem definition is closest to ours is~\cite{DBLP:conf/kbse/NelsonDGK19}, which also uses transition systems.  The authors define a configuration as an initial state of a transition system, which is very similar to our notion of configuration variables.

Constraint solving has been explored in various ways for automating system configuration. Efforts have been made to design declarative, constraint-based, object-oriented languages and policy-based tools to configure systems as well as to validate configurations~\cite{Hewson2013ConstraintbasedSF,DBLP:journals/jnsm/NarainLMK08,Cauldron,Hewsonphdthesis}.
Early approaches were based on constraint satisfaction and constraint logic programming~\cite{Sabin1996ConfigurationAC,SHARMA1998255,tiihonen2013wecotin}.
More recent approaches utilize SAT and SMT solvers~\cite{DBLP:journals/jnsm/NarainLMK08,DBLP:journals/todaes/PeterG15,DBLP:conf/cade/MichelHGH12}, and counterexample-guided inductive synthesis and relational model finding~\cite{DBLP:conf/kbse/NelsonDGK19,Wagner2020WhereTB} for dynamic configuration. However, the way these approaches reduce configuration problems to constraint satisfaction problems is significantly different from our approach using input/output examples and unrolling.  

More significantly, our work differs in its use of modularity and optimization to improve scalability and understandability.
Some automated configuration efforts do employ optimization (e.g., ~\cite{DBLP:conf/lisa/HewsonAG12}), but with a different goal, namely to configure a system in a way that maximizes its performance.

\section{Conclusion}
\label{section:conclusion}
We proposed a new approach for automatically configuring systems representable as transition systems.  Key contributions of our approach include its ability to leverage modularity and its use of optimization. Optimal configurations are more human-understandable, and both modularity and optimization can improve scalability.  We demonstrated these claims with a case study using a CGRA memory tile.

Future directions for this work include applying it to a wider variety of designs, exploring modularity for more sophisticated theories, and finding provably correct configurations for applications with repeating input/output patterns.

  \section*{Acknowledgments}
%\fi

This work was funded in part by the Stanford Agile Hardware Center and by the Defence Advanced Research Projects Agency under grant number FA8650-18-2-7854.

% trigger a \newpage just before the given reference
% number - used to balance the columns on the last page
% adjust value as needed - may need to be readjusted if
% the document is modified later
%\IEEEtriggeratref{8}
% The "triggered" command can be changed if desired:
%\IEEEtriggercmd{\enlargethispage{-5in}}

% references section

% can use a bibliography generated by BibTeX as a .bbl file
% BibTeX documentation can be easily obtained at:
% http://mirror.ctan.org/biblio/bibtex/contrib/doc/
% The IEEEtran BibTeX style support page is at:
% http://www.michaelshell.org/tex/ieeetran/bibtex/
%\bibliographystyle{IEEEtran}
% argument is your BibTeX string definitions and bibliography database(s)
%\bibliography{IEEEabrv,../bib/paper}
%
% <OR> manually copy in the resultant .bbl file
% set second argument of \begin to the number of references
% (used to reserve space for the reference number labels box)

\bibliographystyle{IEEEtran}
\bibliography{main}

\section{Appendix}
\subsection{Appendix A}
Theorem~\ref{Theorem} on the soundness of decomposition uses the $\textsc{GetAbduct}$ function, which generates an abduct of a formula. Exploring different implementations of $\textsc{GetAbduct}$ is an interesting future research direction. Here, we note one sufficient condition for using the simple scheme discussed in  Section~\ref{section:framework}, namely if the theory $\T$ is \emph{complete}.

First, recall that two $\Sigma$-interpretations are \emph{elementarily equivalent} if they satisfy exactly the same closed $\Sigma$-formulas. A theory $\mathcal{T}$ is said to be \emph{complete} if for any closed $\Sigma$-formula $\phi$, either $\phi$ is unsatisfiable or $\neg\phi$ is unsatisfiable. A well-known property of complete theories is that all of their interpretations are elementarily equivalent.

\begin{theorem} If $\mathcal{T}$ is a complete theory, $\phi$ is a formula, $\Model$ is a $\T$-interpretation satisfying $\phi$, and $\psi$ is the formula that assigns the free variables in $\phi$ to their model values from $\Model$, then $\psi$ is an abduct of $\phi$.
\end{theorem}

\begin{proof}
Let $\mathit{free}(\phi):=\{v_1,\dots,v_m\}$ be the set of free variables in $\phi$, and let $s := \Model^{\mathit{free}(\phi)}$. Then, $\psi$ is a conjunction of variable-value qualities from $s$. Consider a closed formula $\phi^c := \exists \, v_1,\dots,v_m.\: (\phi \wedge \psi)$. Since $\Model = \Model[s]$, $\Model[s] \models \phi$. Also, $\Model[s] \models \psi$ by construction. Consequently, $\Model[s] \models \phi \wedge \psi$ and, therefore, $\Model \models \phi^c$. Let $\Model'$ be an interpretation such that $\Model' \models \psi$. Both $\Model$ and $\Model'$ are models of $\mathcal{T}$. $\mathcal{T}$ is complete. Thus, $\Model$ and $\Model'$ are elementarily equivalent and $\Model' \models \phi^c$. Then, $\Model'[s'] \models \phi \wedge \psi$ for some assignment $s'$ over $\{v_1,\dots,v_m\}$. However,  $\Model'[s'] \models \psi$ if and only if $s' = s$. Thus, $\Model'[s'] = \Model'$ and $\Model' \models \phi$. 
\end{proof}

\newpage

\subsection{Appendix B}
This appendix presents more experimental results. We include charts for each studied application separately, for image sizes 16x16 to 60x60. For these experiments, the timeout was set to 1200 seconds for first configurations and 4000 seconds for optimal configurations. Memory limits are 16 GB and 100 GB for the first and the optimal configurations, respectively.

\begin{figure*}
\centering
\begin{subfigure}{.45\textwidth}
  \centering
  \includegraphics[width=.34\paperwidth]{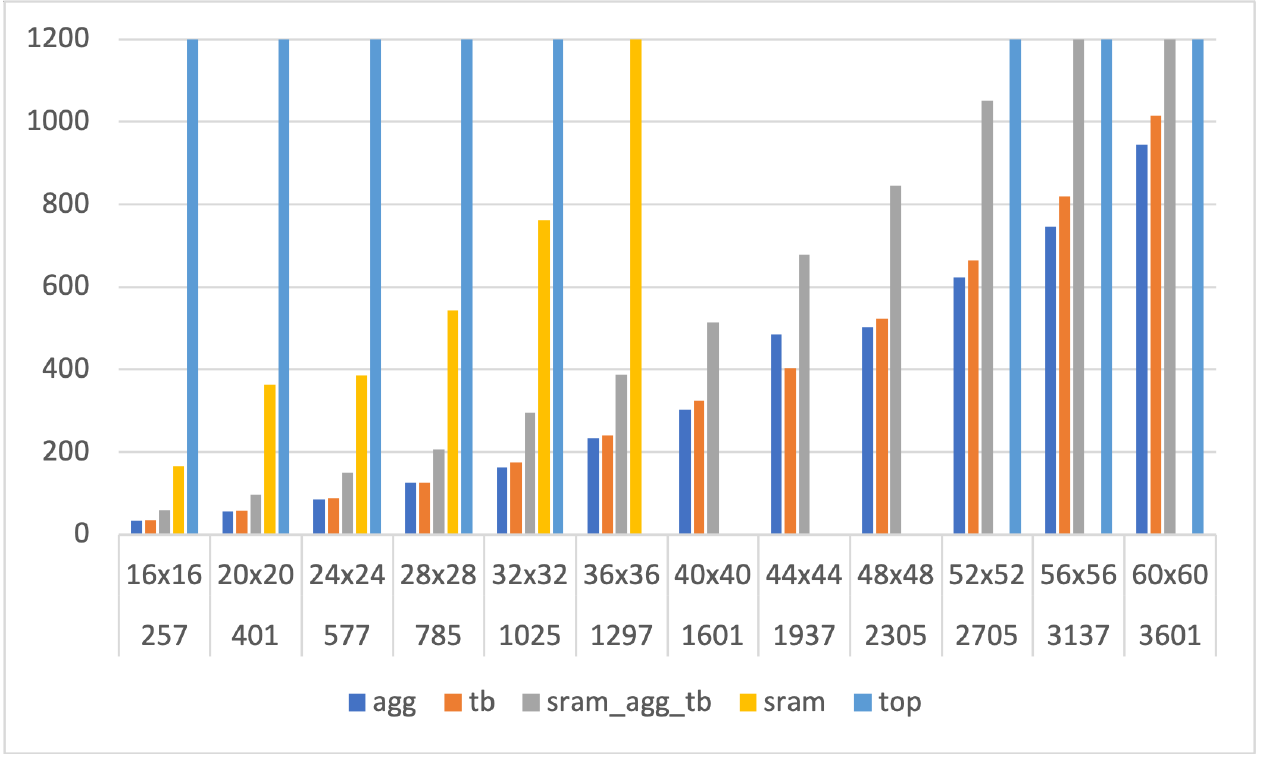}
  \caption{Identity Stream First Configuration.}
  \label{fig:idbest}
\end{subfigure}
\begin{subfigure}{.45\textwidth}
  \centering
  \includegraphics[width=.34\paperwidth ]{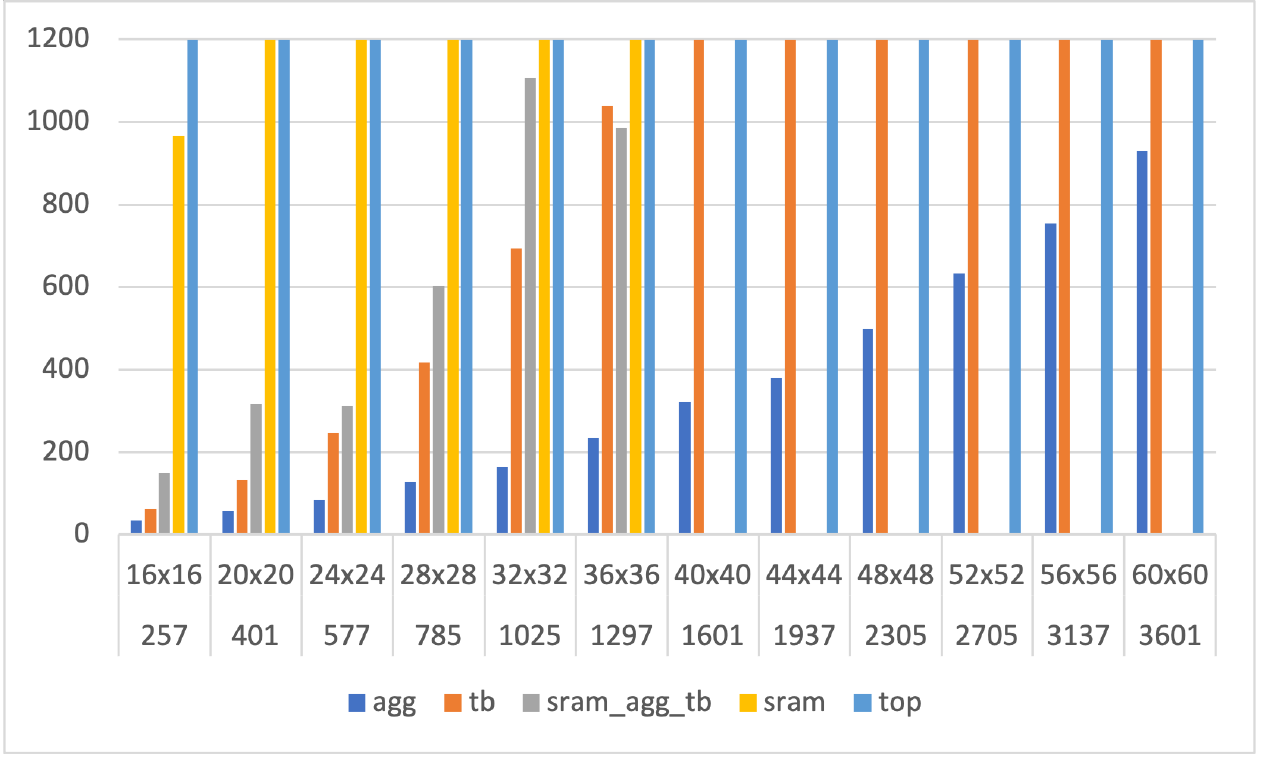}
  \caption{3x3 Convolution First Configuration.}
  \label{fig:convbest}
\end{subfigure}
\begin{subfigure}{.45\textwidth}
  \centering
  \includegraphics[width=.34\paperwidth ]{CascadeConvOpt1.pdf}
  \caption{Cascade (conv) First Configuration.}
  \label{fig:cascconvbest}
\end{subfigure}
\begin{subfigure}{.45\textwidth}
  \centering
  \includegraphics[width=.34\paperwidth ]{CascadeHwOpt1.pdf}
  \caption{Cascade (hw) First Configuration.}
  \label{fig:caschwbest}
\end{subfigure}
\begin{subfigure}{.45\textwidth}
  \centering
  \includegraphics[width=.34\paperwidth ]{HarrisCimOpt1.pdf}
  \caption{Harris (cim) First Configuration.}
  \label{fig:Harriscimbest}
\end{subfigure}
\begin{subfigure}{.45\textwidth}
  \centering
  \includegraphics[width=.34\paperwidth ]{HarrisLxxOpt1.pdf}
  \caption{Harris (lxx) First Configuration.}
  \label{fig:Harrislxxbest1}
\end{subfigure}
\begin{subfigure}{.45\textwidth}
  \centering
  \includegraphics[width=.34\paperwidth ]{HarrisLxyOpt1.pdf}
  \caption{Harris (lxy) First Configuration.}
  \label{fig:Harrislxxbest}
\end{subfigure}
\begin{subfigure}{.45\textwidth}
  \centering
  \includegraphics[width=.34\paperwidth ]{HarrisLyyOpt1.pdf}
  \caption{Harris (lyy) First Configuration.}
  \label{fig:Harrislxxbest}
\end{subfigure}
\begin{subfigure}{.45\textwidth}
  \centering
  \includegraphics[width=.34\paperwidth ]{HarrisPadOpt1.pdf}
  \caption{Harris (pad) First Configuration.}
  \label{fig:Harrislxxbl}
\end{subfigure}
\end{figure*}

\begin{figure*}[t]
\centering
\begin{subfigure}{.45\textwidth}
  \centering
  \includegraphics[width=.34\paperwidth ]{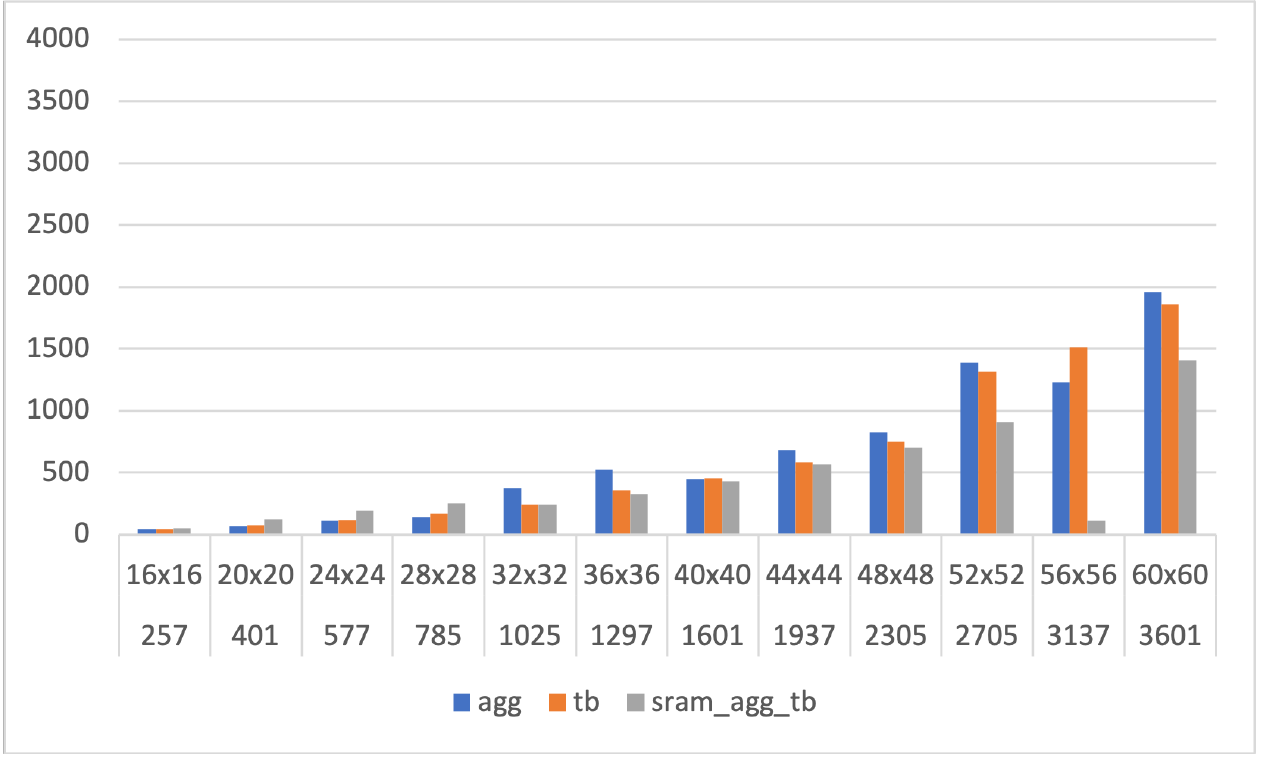}
  \caption{Identity Stream Optimal Configuration.}
  \label{fig:idopt4}
\end{subfigure}%
\begin{subfigure}{.45\textwidth}
  \centering
  \includegraphics[width=.34\paperwidth ]{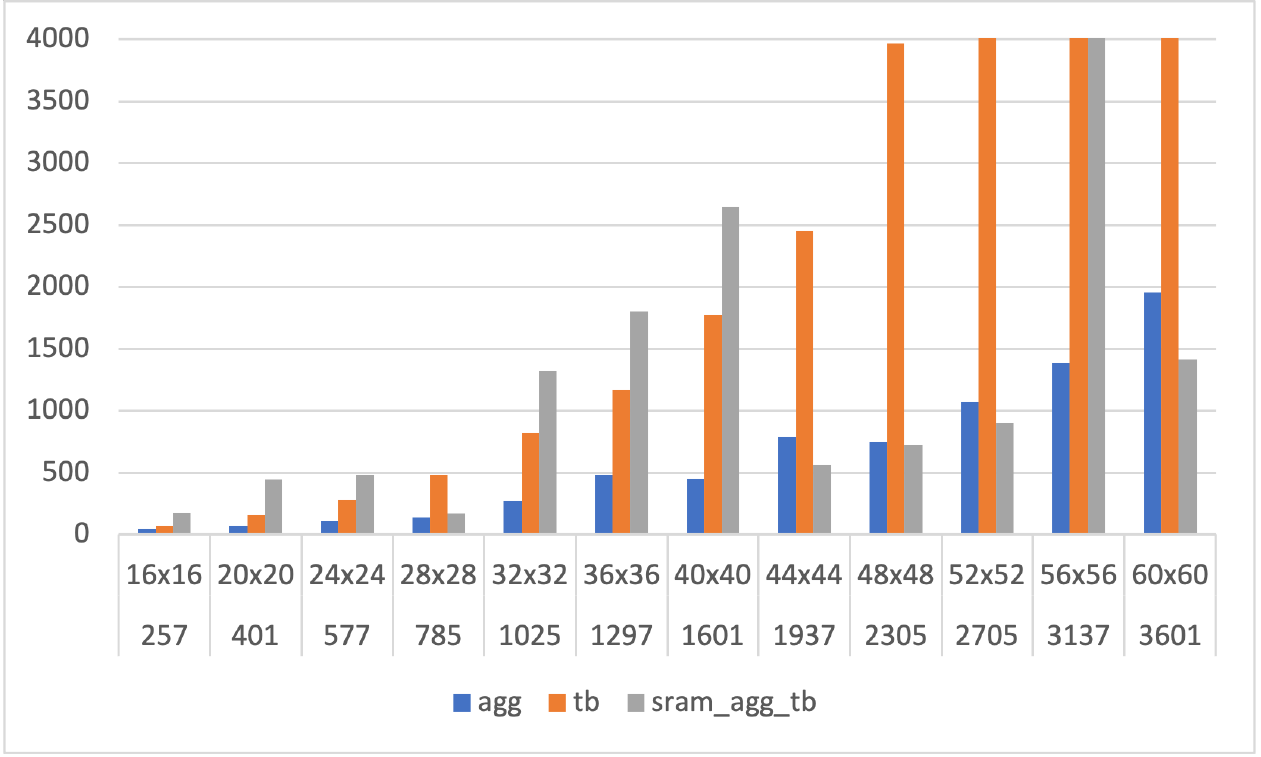}
  \caption{3x3 Convolution Optimal Configuration.}
  \label{fig:convopt4}
\end{subfigure}
\begin{subfigure}{.45\textwidth}
  \centering
  \includegraphics[width=.34\paperwidth ]{CascadeConvOpt4.pdf}
  \caption{Cascade (conv) Optimal Configuration.}
  \label{fig:cascconvopt4}
\end{subfigure}%
\begin{subfigure}{.45\textwidth}
  \centering
  \includegraphics[width=.34\paperwidth ]{CascadeHwOpt4.pdf}
  \caption{Cascade (hw) Optimal Configuration.}
  \label{fig:caschwopt4}
\end{subfigure}
%\caption{Cascade application}
\begin{subfigure}{.45\textwidth}
  \centering
  \includegraphics[width=.34\paperwidth ]{HarrisCimOpt4.pdf}
  \caption{Harris (cim)  Optimal Configuration.}
  \label{fig:Harriscimopt4}
\end{subfigure}%
\begin{subfigure}{.45\textwidth}
  \centering
  \includegraphics[width=.34\paperwidth ]{HarrisLxxOpt4.pdf}
  \caption{Harris (lxx)  Optimal Configuration.}
  \label{fig:Harrislxxopt4}
\end{subfigure}%

\begin{subfigure}{.45\textwidth}
  \centering
  \includegraphics[width=.34\paperwidth ]{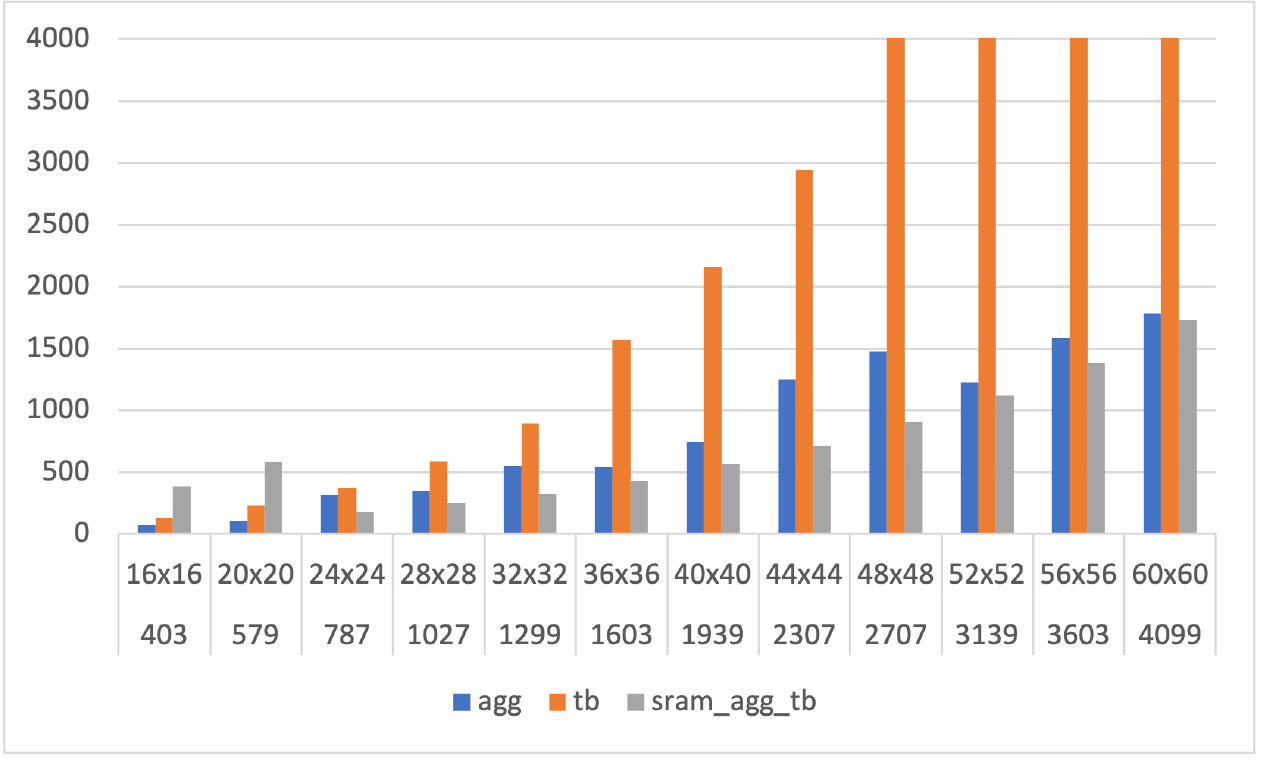}
  \caption{Harris (lxy) Optimal Configuration.}
  \label{fig:Harrislyyopt4}
\end{subfigure}%
\begin{subfigure}{.45\textwidth}
  \centering
  \includegraphics[width=.34\paperwidth ]{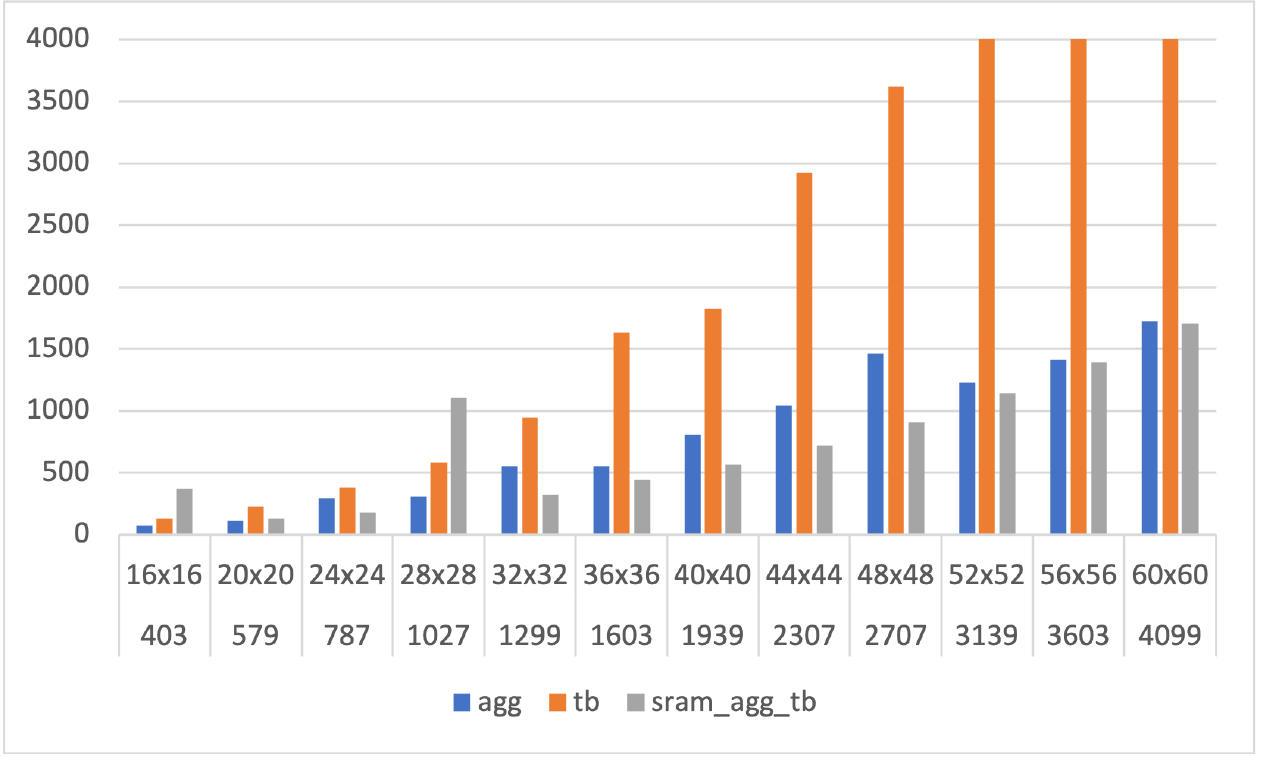}
  \caption{Harris (lyy) Optimal Configuration.}
  \label{fig:Harrispadopt4}
\end{subfigure}
\begin{subfigure}{.45\textwidth}
  \centering
  \includegraphics[width=.34\paperwidth ]{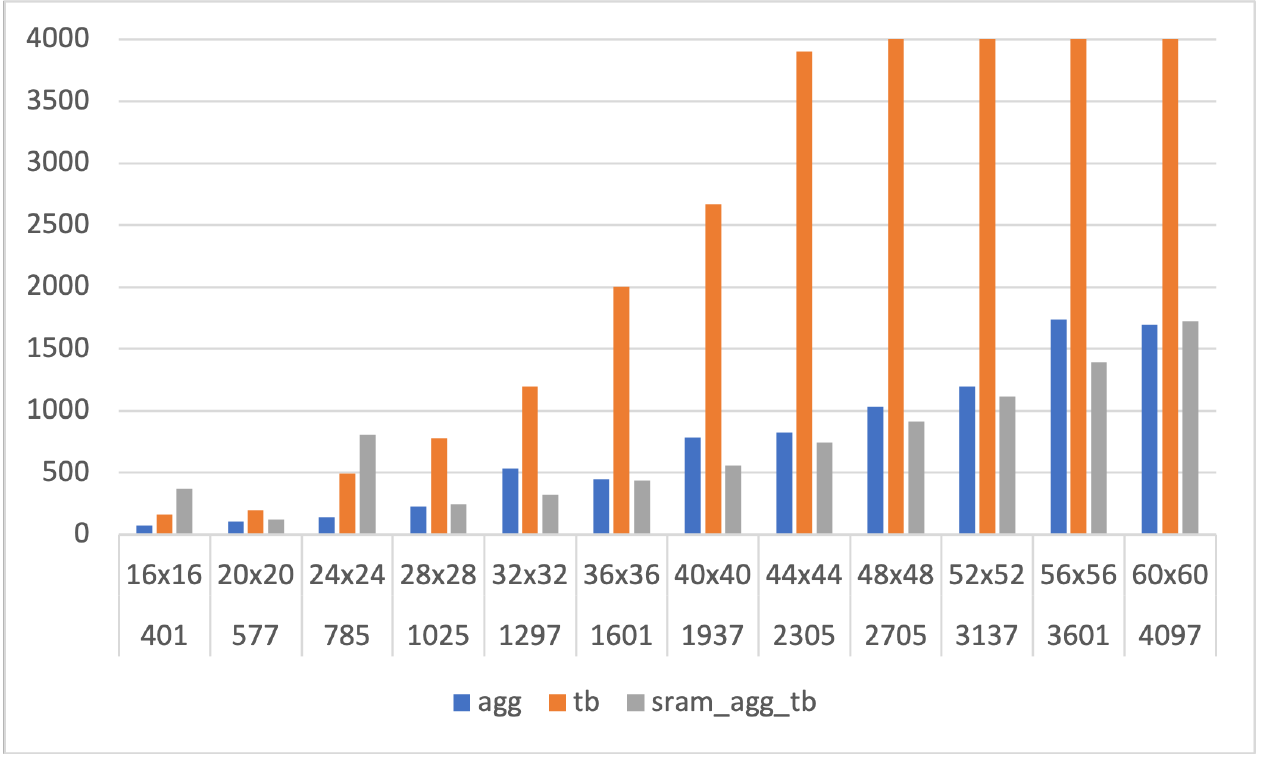}
  \caption{Harris (pad) Optimal Configuration.}
  \label{fig:Harrislxyopt4}
\end{subfigure}
\end{figure*}

% that's all folks
\end{document}